\documentclass[sigplan,screen,sigconf,nonacm]{acmart}

\usepackage{pervasives}

\author{Rachit Nigam*}
\affiliation{
  \institution{Cornell University}
  \country{USA}
}
\author{Samuel Thomas*}
\affiliation{
  \institution{Cornell University}
  \country{USA}
}
\author{Zhijing Li}
\affiliation{
  \institution{Cornell University}
  \country{USA}
}
\author{Adrian Sampson}
\affiliation{
  \institution{Cornell University}
  \country{USA}
}

\begin{document}

\title{%
  \papertitle{}
}

\begin{abstract}
We present \sys{}, a new intermediate language (IL) for compiling
high-level programs into hardware designs.
\sys{} combines a hardware-like structural language with a software-like
control flow representation with loops and conditionals.
This split representation enables a new class of hardware-focused optimizations
that require both structural and control flow information which are crucial
for high-level programming models for hardware design.
The \sys{} compiler lowers control flow constructs using finite-state machines
and generates synthesizable hardware descriptions.

We have implemented \sys{} in an optimizing compiler that translates
high-level programs to hardware.
We demonstrate \sys{} using two DSL-to-RTL compilers,
a systolic array generator and
one for a recent imperative accelerator language, and compare them to
equivalent designs generated using high-level synthesis (HLS).
The systolic arrays are \systolicArrayCycles faster and \systolicArrayLut larger on average than HLS
implementations, and the HLS-like imperative language compiler is within a few
factors of a highly optimized commercial HLS toolchain.
We also describe three optimizations implemented in the \sys{} compiler.

\end{abstract}

\maketitle

\section{Introduction}

\begin{figure*}
  \centering
  \begin{subfigure}[b]{0.40\linewidth}
    \centering
    \includegraphics[width=\linewidth]{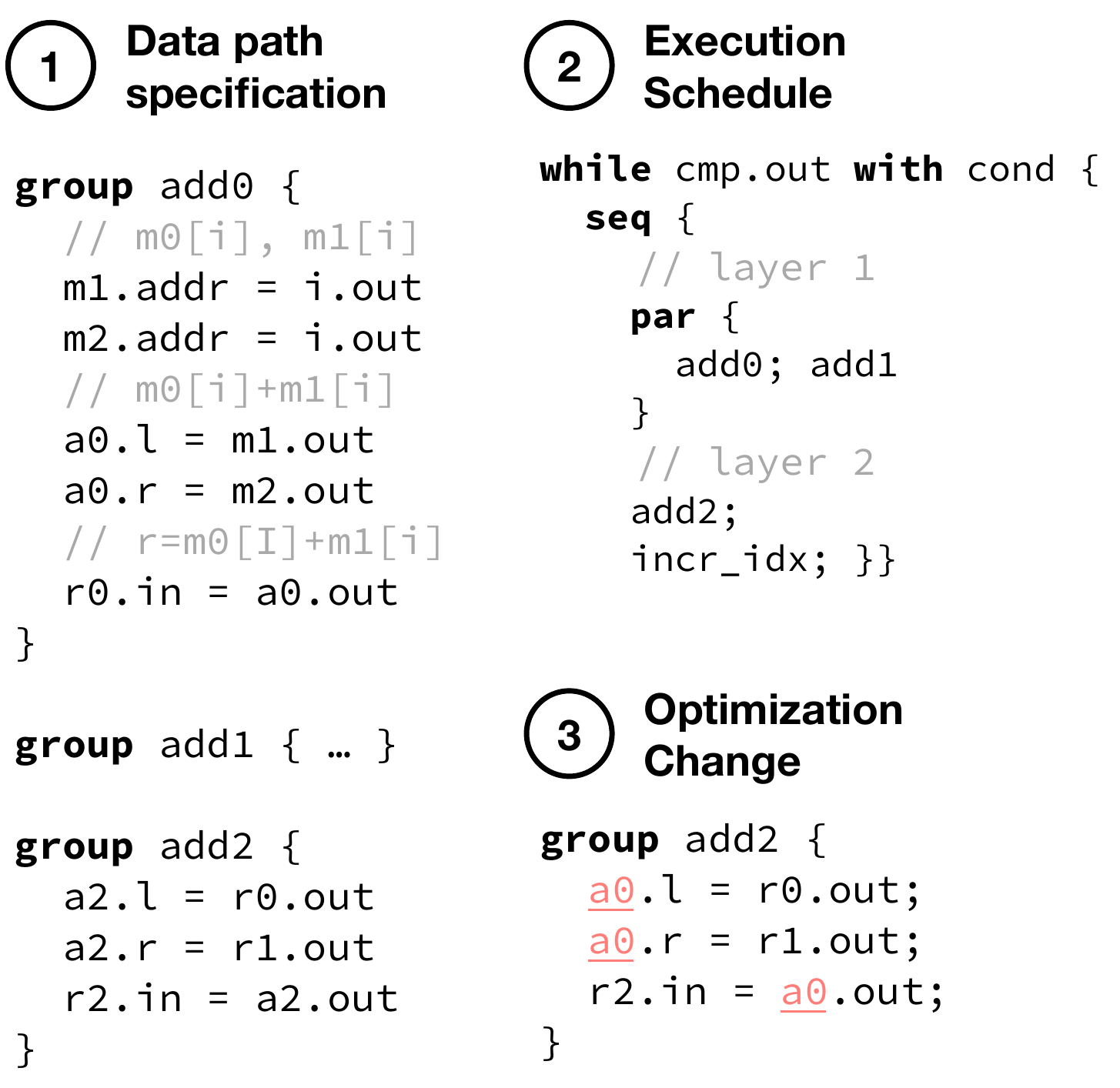}
    \caption{\sys{} program. Groups \code|incr_idx| and \code|cond| elided.}
    \label{fig:reduction:code}
  \end{subfigure}
  \hfill
  \begin{subfigure}[b]{0.32\linewidth}
    \centering
    \includegraphics[width=\linewidth]{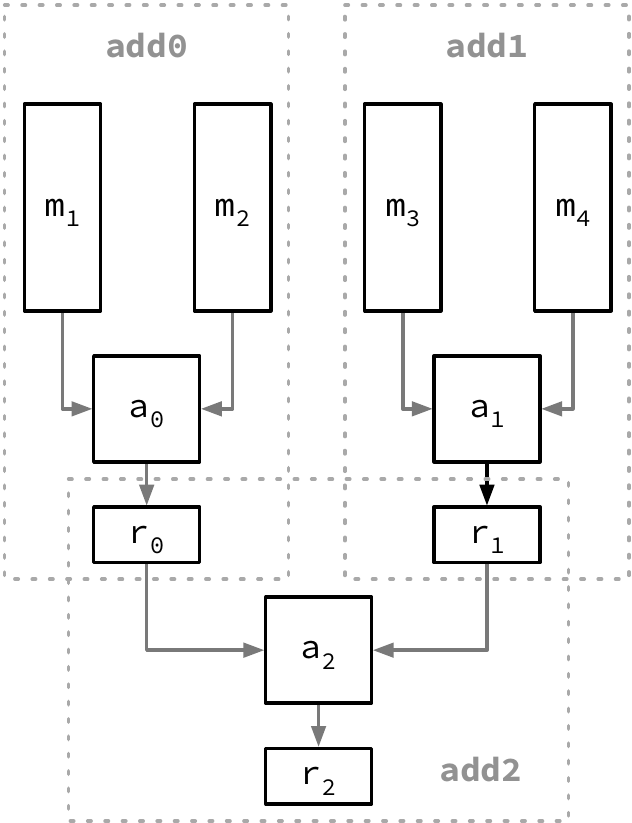}
    \caption{Initial architecture (groups marked).}
    \label{fig:reduction:orig}
  \end{subfigure}
  \hfill
  \begin{subfigure}[b]{0.24\linewidth}
    \centering
    \includegraphics[width=\linewidth]{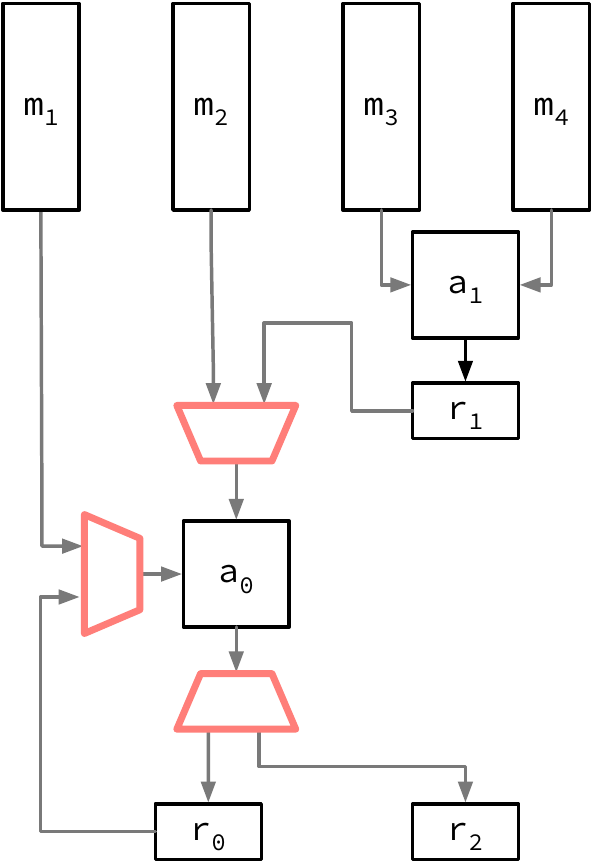}
    \caption{Optimized architecture.}
    \label{fig:reduction:opt}
  \end{subfigure}
  \caption{\sys{} describes the reduction tree using its split representation.
The execution schedule makes the control flow explicit while encapsulate connections between hardware modules.
Done signals (\cref{sec:lang:groups-and-control}) elided from group definitions.}
  \label{fig:reduction}
\end{figure*}

\blfootnote{\emph{* Equally contributing authors.}}
Hardware design is a language problem.
While custom hardware accelerators are economically justified in a post
Moore's law era, we have yet to see widespread adoption.
Even though reconfigurable architectures, such as field programmable gate arrays (FPGAs), make it easy to
deploy accelerators, the tooling and languages inhibit ubiquitous use.
Hardware description languages (HDLs) operate at the level of gates, wires, and
clock cycles; while this level of abstraction is useful for designing
high-end processors, it is inappropriate for the rapid design of
computational accelerators.

To liberate hardware design from these low-level abstractions,
researchers have proposed several compilers for high-level specification languages.
The traditional approach is
high-level synthesis (HLS): to compile legacy software languages such as C,
\cxx, or OpenCL to HDLs~\cite{vivadohls2017,
catapulthls, legup, autopilot, intelhls}.
However, such languages are a poor fit for generating hardware---they reflect
pointer-based, sequential, von~Neumann models of computation.
The hardware
they seek to generate is pervasively parallel, without a unified address space,
and free from program counters.

The cavernous semantic gap between \cxx and HDLs motivates a more
domain-specific approach. A new wave of hardware languages and compilers
focus on a
specific application category~\cite{halide-hls, p4fpga},
on a specific architecture style~\cite{aetherling},
or on lifting hardware-level concerns into a restricted
imperative language~\cite{dahlia, spatial}.
These narrower languages sacrifice
the familiarity and backwards compatibility of traditional HLS to simplify
compilation, generate better hardware, and avoid the uncanny valley of
inconsistent software-like semantics. They can focus on providing high-level
abstractions that concisely capture the parallelism of the application
domain.

DSL-to-hardware compilers, however, remain substantial feats of engineering. The
compiler developer needs not only to conceive of a high-level architecture; they
must also design a data path and a control path to implement the execution
strategy and perform architectural optimizations~\cite{aetherling, spatial}.
Each such compiler re-engineers a new intermediate language (IL) to encode the
high-level semantics of the input language while exposing
architectural information to perform optimizations. A shared IL, along with a
compiler infrastructure that implements useful optimizations and analyses, will
let compiler engineers design new hardware DSLs and quickly get competitive
hardware designs.

We propose \sys{}, a new intermediate language for compiling DSLs to
hardware.
\sys{} combines a software-like imperative sub-language, which explicitly
represents the control flow of a design, with a structural language, which
instantiates hardware modules and describes connections between them.
Frontend compilers can specify architectural details using the structural
sub-language and rely on the high-level control language to encode a DSL's
semantics.
The \sys{} compiler optimizes these programs, generates control logic, and
emits synthesizable RTL.

The contributions of this paper are:
\begin{itemize}
\item
\sys{}, an intermediate language for compiling DSLs to hardware that uses
a split representation combining a high-level control flow language with a
hardware-like structural language.

\item
An \href{https://github.com/cucapra/calyx}{open-source} pass-based
compiler for analyzing, optimizing, and lowering \sys{} programs to
synthesizable RTL.

\item
The implementation of two compilers that target \sys{}:
(1) a PE-parametric systolic array generator that encodes the data movement and
computation schedule using \sys{}'s control language, and
(2) Dahlia~\cite{dahlia}, a general-purpose programming language for
accelerator design which has a preexisting backend targeting HLS toolchains.

\item
Three optimizations implemented within the \sys{} compiler: resource sharing,
live-range-based register-sharing, and a pass to infer cycle latencies.
\end{itemize}

\section{Overview by Example}

This section introduces \sys{} by using it to implement a parallel
\emph{reduction tree}.
A reduction tree applies an operator to many inputs to produce a single output.
\Cref{fig:reduction:orig} shows a small summation tree on four inputs.
The operators within a tree level run in parallel to produce the inputs to the
next level.
Unlike hardware description languages (HDLs) or high-level synthesis (HLS),
\sys{} programs are meant to be generated by compiler frontends.
We show that with \sys{}'s control language, compilers can encode the semantics of high-level languages while
producing programs amenable to hardware optimization.

\subsection{Reduction Tree in \sys{}}

\Cref{fig:reduction:code} shows a \sys{} program fragment that implements a parallel
reduction tree that computes $(m_0 + m_1) + (m_2 + m_3)$.
The program uses \emph{groups} to specify the data path \circled{\sf\textbf1}.
Groups encapsulate hardware connections that implement an action.
For example, the group \code{add0} uses the hardware adder \code{a0} to compute
the sum of the first two inputs and save the result in a register \code{r0}.
The assignments used inside groups correspond to \emph{non-blocking assignments}
in RTL languages---updates to the left hand side of an assignment are
immediately propagated to the right hand side.
In this way, each group encapsulates a data flow graph.

To compute the reduction, we need to schedule the execution of the layers.
We want to execute the layers sequentially
and to run the adders inside a tree layer in parallel.
The \sys{} program specifies the reduction tree's schedule using a separate \emph{control} language \circled{\sf\textbf2}.
The control language uses group names to activate hardware connections.
Unlike groups, control statements have no direct hardware analog---instead,
they resemble a small imperative program with explicit parallelism.
The schedule iterates over the memories using a \code{while} statement
and sequences the execution of the layers using the \code|seq| operator.
The \code|par| operator specifies that the adders in the first layer will be
executed in parallel.
Finally, the loop body uses the group \code{incr_idx} to increment the index
into the memories.

\Cref{fig:reduction:orig} shows the high-level architecture generated from the
\sys{} program and marks the connections that correspond to the groups.
The figure elides the control circuitry generated to implement the schedule.

\subsection{Optimizing Accelerator Designs}

High-level specifications of accelerators encode a treasure trove of control
flow information that is lost when lowering to a register-transfer level (RTL)
language.
Compilers for such programming models need a stable intermediate language (IL)
to capture and use such information.
However, RTL is ill-suited for this task.

RTL languages do not distinguish between control flow and data flow because
they implement both using the same structural constructs.
For example, in order to sequence two operations, an RTL program must implement
a state machine to track the current state.
Such a state machine is implemented using registers and adders which are
indistinguishable from registers and adders used to implement the program's
data flow.
This conflation means that a compiler cannot automatically extract and
transform the control flow of an arbitrary RTL program.

Consider an optimization that reuses existing circuitry to perform temporally
disjoint computations.
For example, our reduction tree uses adders \code|a0| and \code|a2|
in two different stages and never overlaps their execution.
Therefore, it would be safe to transform the program to share a single adder
for both the stages.
Implementing this optimization in RTL, however, is difficult
because the structural implementation of a state machine obscures the program's
control flow.
To determine that the two adders run at different times, an analysis would need
to reverse-engineer the execution schedule from the state machine
implementation.
Furthermore, transforming an RTL program would require pervasive changes.
\Cref{fig:reduction:opt} shows the optimized architecture.
The transformation requires carefully rewiring the input and output signals
for \code|a0| through multiplexers.

In contrast, a \sys{} program makes the control flow explicit and enables
straightforward transformation.
Given the execution schedule of our \sys{} program, it is clear that the
groups \code|add0| and \code|add2| do not execute simultaneously since they
are scheduled using the \code|seq| operator.
\Cref{fig:reduction:code} \circled{\sf\textbf3} shows the only change required
to implement this optimization.
The \sys{} program simply renames the uses of \code|a2| in group \code|add2|
with \code|a0| and the compiler correctly generates the additional multiplexers
and control signals to share the adder.

\subsection{Structure and Control}

\sys{} is neither a software IL nor a hardware IL.
Software ILs, such as LLVM~\cite{llvm}, focus on providing a uniform representation
of the control flow and data flow of a program.
They do not explicitly represent structural facts, such as the mapping of logical
adds onto physical adders.
On the other hand, hardware ILs focus on a purely structural representation
with explicit use of gates, wires, and clocks while conflating data flow with
control signals.
By marrying structure and control,
\sys{} provides access to both structural and control flow facts to enable a
new class of optimizations that cannot be captured by either style of ILs.

\section{The \sys{} Intermediate Language}
\label{sec:lang}

The \sys{} infrastructure's focal point is its program representation.
The \sys{} IL aims to represent domain-specific accelerator designs throughout the entire lifetime of a hardware generation pipeline:
generation from a language frontend,
optimization and lowering,
and implementation in a hardware description language.
This section describes the \sys{} IL;
the following sections show how to generate, lower and optimize the IL.

\subsection{Components}

\sys{} programs consist of \emph{components} which encapsulate
hardware structures and define an execution schedule to orchestrate their
behavior:
\begin{lstlisting}
component $\textit{name}$($\textit{inputs}$) -> ($\textit{outputs}$) {
  cells { ... }
  wires { ... }
  control { ... }
}
\end{lstlisting}
The body includes hardware-like structural listings of \emph{cells} and
\emph{wires} (\cref{sec:lang:cells-and-wires})
and software-like \emph{control} code (\cref{sec:lang:groups-and-control}).
The input and output ports form the interface to the component and define
their size in bits.
For example, a component defining a 32-bit integer adder uses these ports:
\begin{lstlisting}
component adder(lhs: 32, rhs: 32) -> (sum: 32)
\end{lstlisting}
Ports in \sys{} are \emph{untyped}---they can hold any value of a given width.
\sys{} leaves type-based reasoning to the language frontend.

\subsection{Cells and Wires}
\label{sec:lang:cells-and-wires}

\sys{} programs explicitly instantiate components and define the connections
between them in a way that closely resembles RTL languages.
This low-level of detail gives frontends precise control over fine-grained
architectural choices when needed and lets \sys{} lower programs to
synthesizable RTL.

The cells section instantiates components:
\\
\begin{minipage}{1.0\linewidth}
\begin{lstlisting}
cells {
  a_reg = std_reg(32); // 32-bit register
  add = std_add(32);   // 32-bit adder
}
\end{lstlisting}
\end{minipage}
\\
This example instantiates a register and an adder that operate on 32-bit
values using the \code|std_reg| and \code|std_add| components.
The wires section defines \emph{assignments} between the ports of components:
\begin{lstlisting}
wires {
  add.left = a_reg.out;
  add.right = a_reg.out;
}
\end{lstlisting}
These \emph{assignments} connect the \code{out} port of the register to the
two input ports of the adder.
The connections are \emph{non-blocking}: updates to \code{a_reg.out} are
immediately visible to \code{add.left}.
This closely resembles non-blocking assignments in RTL languages.

Wire assignments can specify more complex dataflow policies by using
\emph{guarded assignments}:
\begin{lstlisting}
add.left = cmp.out ? a_reg.out;
add.left = !cmp.out ? b_reg.out;
\end{lstlisting}
The guarded assignments to the \code{left} port of the \code{add} component
use the value of \code{cmp.out} to determine the assignment to activate.
Guards are built with ports and a simple language of boolean connectives.

Like its RTL counterparts, \sys{} requires that each port have a \emph{unique
driver}---activating multiple assignments in the same cycle results
in undefined behavior.
This requirement also differentiates \sys{}'s guarded assignments from
Bluespec's atomic rules~\cite{nikhil:bluespec}.
While Bluespec resolves conflicting assignments by generating scheduling logic
to dynamically abort them, \sys{} does not.
Being an intermediate language, \sys{} trades-off the convenient programming
abstraction for predictable compilation.

Guarded assignments in \sys{} correspond exactly to assignments in RTL
languages.
By themselves, they can encode arbitrary hardware designs, but are less amenable
to analysis and transformation.
The next section describes \sys{}'s two novel constructs that simplify the
specification of a program's structural connections and its execution schedule.

\subsection{Groups and Control}
\label{sec:lang:groups-and-control}

\sys{} uses \emph{groups} to encapsulate assignments.
Inside a group, assignments must obey the same constraints as RTL---unique
drivers for ports, no combinational loops, etc.
However, multiple groups can use the same port:
\begin{lstlisting}
group assign_one { x_reg.in = 1; }
group assign_two { x_reg.in = 2; }
\end{lstlisting}
Both groups unconditionally assign to the same port.
However, since the groups encapsulate the assignments, they are not active by
default and do not violate the unique driver requirement.
In contrast, RTL languages require programmers to reason about all assignments
to a port and weave in control signals to define a unique driver.

The \emph{control program} determines when groups run:
\begin{lstlisting}
control { seq { assign_one; assign_two } }
\end{lstlisting}
The control block uses the \code{seq} (sequence) statement to specify that
\code{assign_one} executes first,
followed by \code{assign_two}.
Since the two groups execute at different times, the two assignments to the
port \code{x_reg.in} do not conflict and \sys{} can generate valid RTL.

While control statements like \code|seq| can pass the control flow of a program
to a group, they have no way to know when to return---groups can encode
arbitrary computations that don't have an obvious done condition.
To signal when it has finished executing, a group use a \code|done| signal:
\begin{lstlisting}
group assign_one {
  x_reg.in = 1;
  assign_one[done] = x_reg.done;
}
\end{lstlisting}
In the above group, we are writing a value to a stateful element \code{x_reg},
and must wait for the element to signal that the write was committed.
The group uses the \code{x_reg.done} port to signal that the group's
computations has finished.

Interface signals, such as a group's done signal, are used by \sys{}
to define a \emph{calling convention} (\Cref{sec:compile:holes}).
A control program passes control flow to a group by setting a group's \code|go|
to $1$ and the group returns control by setting its \code|done| signal to $1$.
Similarly, components use \code|go| and \code|done| interface signals to define
a consistent calling convention.
\sys{}'s interface is \emph{latency-insensitive}; it does not
not reason about the number of cycles needed to execute a computation.
\Cref{sec:compile:static} shows how enriching \sys{} programs with latency
information enables efficient compilation.

\subsection{Control Statements}
\label{sec:lang:control-prims}

\sys{} provides several primitives to encode the schedule of components.
We designed these primitives to capture high-level properties such as branching
and looping, freeing frontends from the need to realize them in control
circuitry.

\paragraph{enable}
Naming a group inside a control statement passes control to the group.

\paragraph{par}
List of control statements that execute once in parallel.
\begin{lstlisting}
par { group_a; seq { group_b; group_c; }; group_d; }
\end{lstlisting}

\paragraph{seq}
List of control statements executed in order.
\begin{lstlisting}
seq { group_a; par { group_b; group_c; }; group_d; }
\end{lstlisting}

\paragraph{if}
Conditionally executes one of the branches.
Specifies a port to use as the $1$-bit conditional value (\emph{port\_name})
and a group (\emph{cond\_group}) to compute the value on the port.
\begin{lstlisting}
if $\textit{port\_name}$ with $\textit{cond\_group}$ {
  $\textit{true\_stmt}$
} else {
  $\textit{false\_stmt}$
}
\end{lstlisting}

\paragraph{while}
The loop statement is similar to the conditional.
It enables \emph{cond\_group} and uses \emph{port\_name} as the conditional
value.
When the value is high, it executes \emph{body\_stmt} and recomputes the
conditional using \emph{cond\_group}.
\begin{lstlisting}
while $\textit{port\_name}$ with $\textit{cond\_group}$ {
  $\textit{body\_stmt}$
}
\end{lstlisting}

\subsection{Attributes}
\label{sec:lang:attributes}

\sys{} programs can use \emph{attributes} to encode frontend and pass-specific
information such as the latency of a group.
Attributes are key-value pairs.
For example, the following group defines an attribute ``latency'' and associates
the value $1$ to it.
\begin{lstlisting}
group foo<"latency"=1> { ... }
\end{lstlisting}

\subsection{Synopsis}

Components are the building blocks of \sys{} programs.
Each component instantiates subcomponents (\emph{cells}) and defines the
connections between them (\emph{wires}).
The \emph{control} program defines the execution schedule by enabling groups.

The design principle behind \sys{} is thus: in general, frontends
generate small groups to perform simple actions, such as incrementing a register
or comparing values, and use the control flow program to schedule them.
However, when frontends have domain-specific knowledge, they can instantiate
complex architectures and encapsulate them using groups.

\section{Compiling \sys{} to Hardware}
\label{sec:compile}

\begin{figure*}
  \centering
  \begin{subfigure}[b]{0.225\linewidth}
    \centering
    \includegraphics[height=\linewidth]{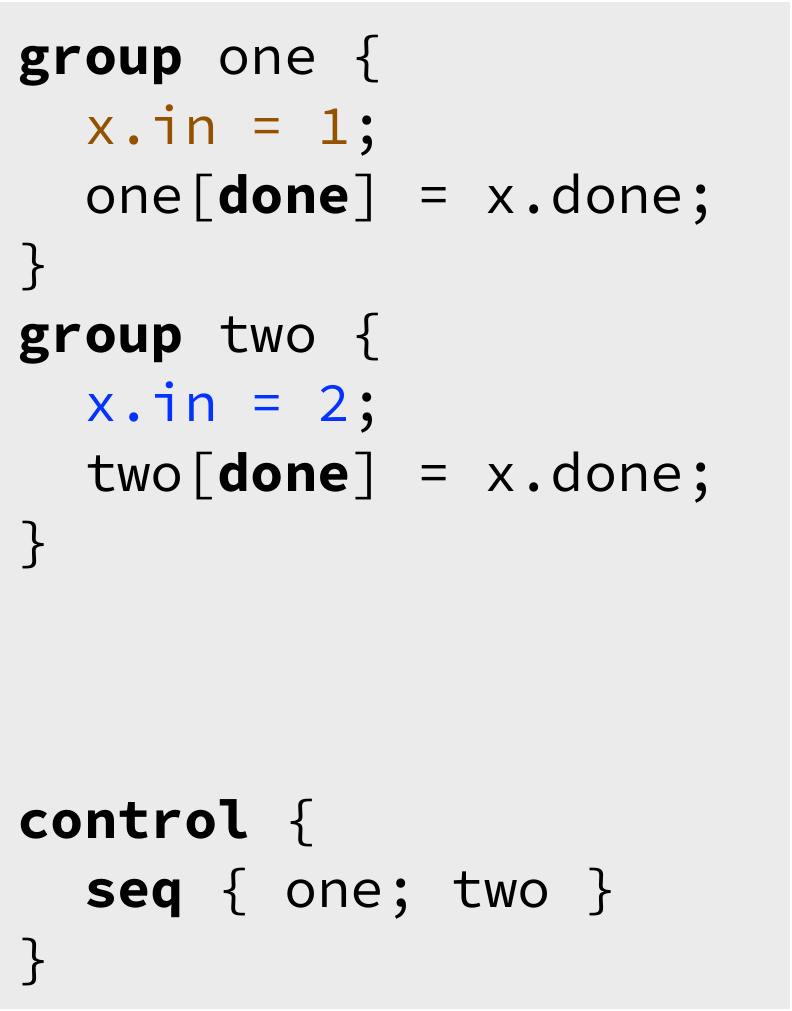}
    \caption{Original program}
    \label{fig:compilation:input}
  \end{subfigure}
  \hspace{-15pt}  
  \begin{subfigure}[b]{0.178\linewidth}
    \centering
    \includegraphics[width=\linewidth]{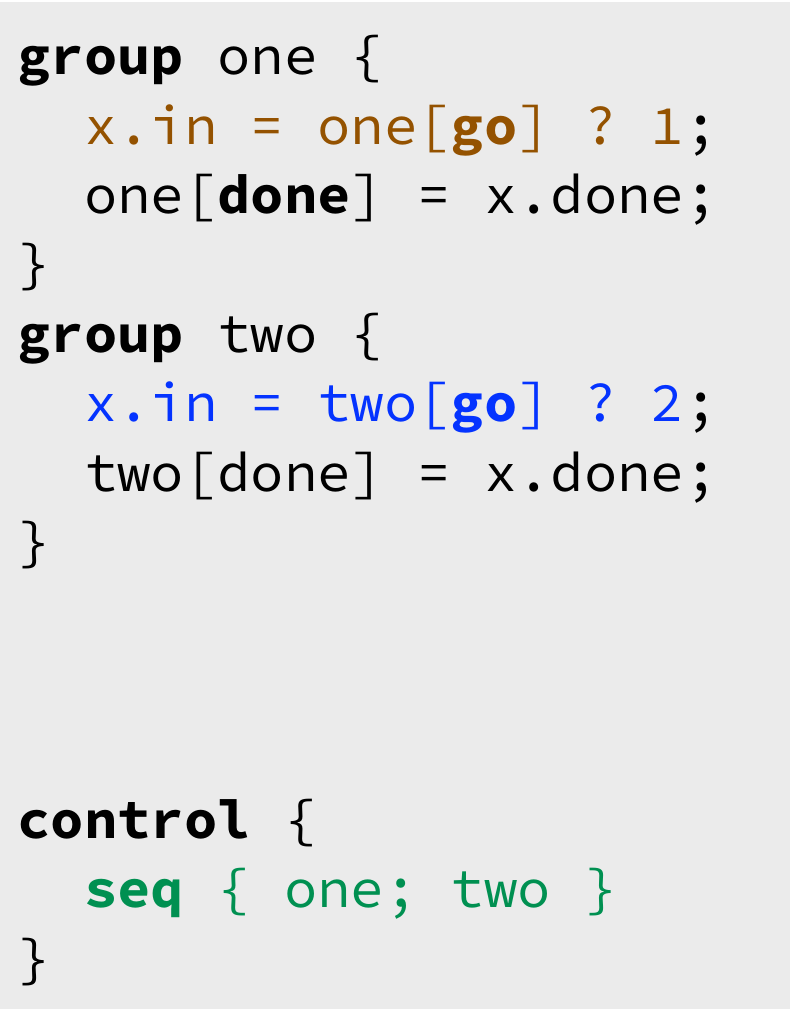}
    \caption{GoInsertion}
    \label{fig:compilation:go-insertion}
  \end{subfigure}
  \begin{subfigure}[b]{0.28\linewidth}
    \centering
    \includegraphics[width=\linewidth]{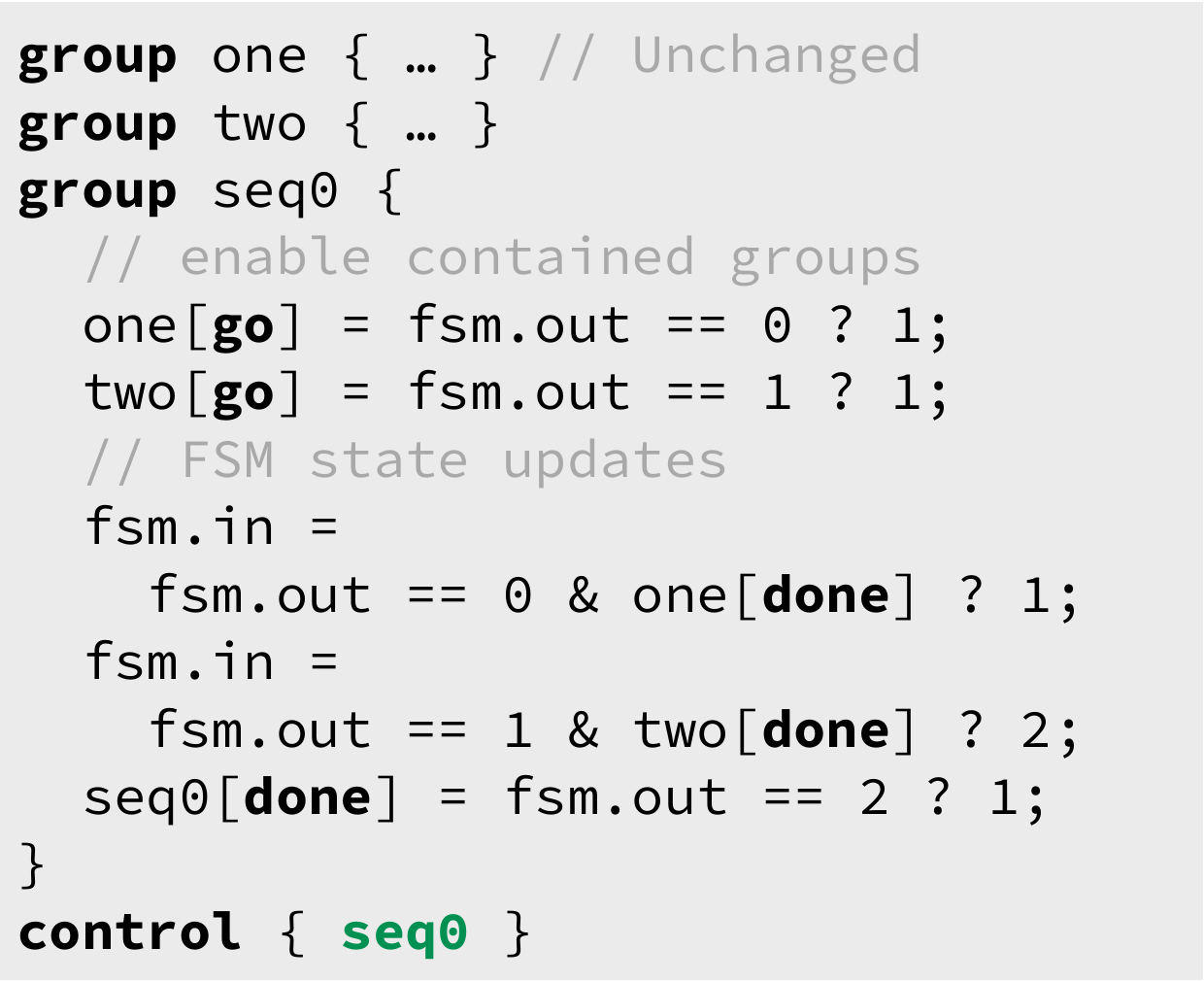}
    \caption{CompileControl}
    \label{fig:compilation:compile-control}
  \end{subfigure}
  \begin{subfigure}[b]{0.295\linewidth}
    \centering
    \includegraphics[width=\linewidth]{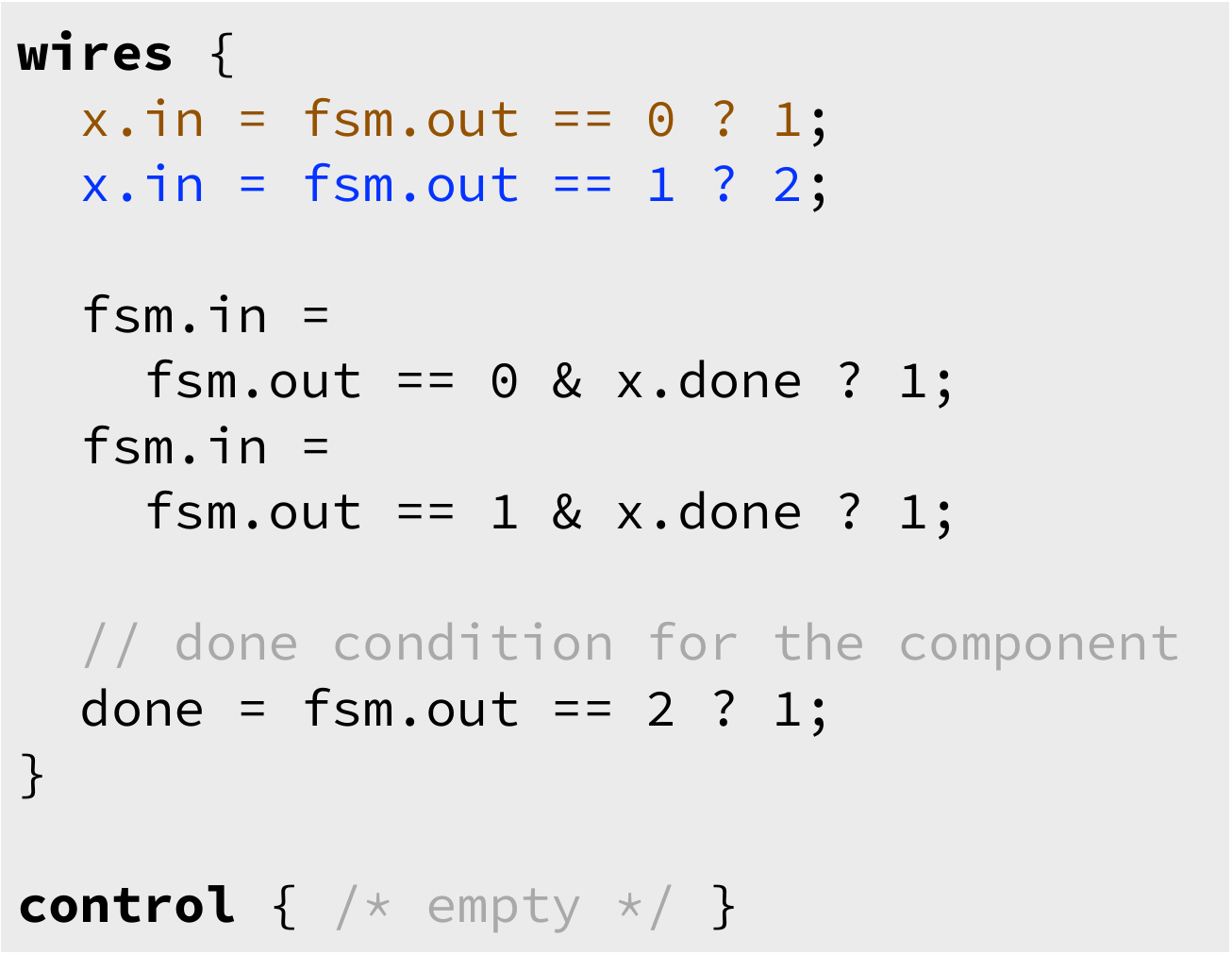}
    \caption{RemoveGroups}
    \label{fig:compilation:remove-holes}
  \end{subfigure}

  \caption{\sys{} realizes the execution schedule by encoding it with structural components. After the \textsc{CompileControl} pass (c), the \code{fsm} register encodes the current state for the \code{seq} statement.}
  \label{fig:compilation}
\end{figure*}

The \sys{} compiler optimizes (\cref{sec:opt}) and lowers \sys{} programs
into synthesizable RTL.
Compilation passes use \emph{interface signals}, which define a calling
convention, to realize a component's execution schedule.
The result is a \sys{} program with a flat list of guarded assignments and
no control statements or groups.
The compiler can then directly translate this flattened form into RTL.
The primary compilation passes are:
\begin{itemize}
\item
\pass{GoInsertion}: Guards all assignments in a group with the group's \code{go}
interface signal.

\item
\pass{CompileControl}: Generates latency-insensitive finite state machines
to structurally realize control operators.

\item
\pass{RemoveGroups}: Inlines uses of interface signals and eliminates all
groups.

\item
\pass{Lower}: Translates control-free \sys{} to RTL.

\item
\pass{Sensitive}: Opportunistically compiles control statements into groups
using latency-sensitive FSMs. Only affects groups with \code|"static"|
attribute.
\end{itemize}
\Cref{fig:compilation} illustrates the main steps.
This section describes the complete compilation process.

\subsection{Calling Convention}
\label{sec:compile:holes}

To realize a \sys{} program's execution schedule, the compiler needs a
mechanism to pass control flow in purely structural programs.
We use a pair of \emph{interface signals} to define this interface:
when a group sets another group's \code{go} signal high, control is passed to
that group and it can enable assignments within it;
when a group sets its own \code{done} signal high, it passes control back.
This interface resembles traditional latency-insensitive hardware
design~\cite{carloni:latency-insensitive}.

Most passes treat interface signals like any other $1$-bit port.
The main compilation passes treat them specially---using them to wire up the
control signals.
The final compilation pass eliminates interface signals by inlining them.

\subsection{Compilation Workflow}
\label{sec:compile:example}

We describe the compilation pipeline by compiling the example \sys{} program in
\cref{fig:compilation:input}.

\paragraph{Inserting \code{go} interface signals}
\sys{}'s semantics requires that assignments within a group are only enabled
when the group executes.
To enforce this requirement, the \pass{GoInsertion} pass inserts the group's
go signal into the guards of the contained assignments.
\Cref{fig:compilation:go-insertion} shows the resulting program:
\code|one[go]| guards assignments in group \code|one| while \code|two[go]|
guards assignments in group \code|two|.
When all groups are eventually removed, these guards will ensure that the
correct set of assignments are active at a given time.

\paragraph{Compiling control using interface signals}
The next step in the compilation process is realizing the control program
using a structural implementation.
Compilation relies on two important properties of \sys{}:
(1) groups can encode arbitrary computations, and
(2) all groups are treated uniformly, regardless of the computation
they perform---a group that increments a register is compiled the same way
as a group that runs a systolic array.

The \pass{CompileControl} pass performs a bottom-up traversal
of the control program and does the following:
(1) for each control statement, such as \code|seq| or \code|while|,
instantiate a new group, called the \emph{compilation group}, to contain all the
structure needed to realize the control statement,
(2) implement the schedule by setting the constituent groups' go and
done signals, and
(3) replace the statement in the control program with the corresponding
\emph{compilation group}.
After this pass, every component's control program is reduced to a single group
enable.

\Cref{fig:compilation:compile-control} shows these transformations.
The pass defines a new group \code|seq0| to encapsulate the structure required
to realize the \code|seq| statement as well as a new register \code|fsm|
to track the current state.
Next, the pass enables the groups contained in the \code|seq| by writing to their
go interface signals and updates the FSM state when the groups set their
done signal high.
The done condition for \code|seq0| is when the FSM reaches its final
state.
Finally, the pass replaces the \code|seq| control statement with the group
\code|seq0|.

\paragraph{Inlining interface signals}
The \pass{RemoveGroups} pass inlines all uses of interface signals and
removes all groups.
It performs three transformations:
\begin{enumerate}
\item
Add new \code{go} and \code{done} \emph{ports} to each component definition
and wire them up to the single group enable in the control program.

\item
Collect all writes to a group's go and done signals and inline them into all
uses of the signals.
If there are multiple writes to a signal, replace the corresponding reads with
a disjunction of the written expressions.
This step eliminates all interface signals from the component.

\item
Remove all groups.
Since all assignments are guarded by expressions that encode the schedule,
it is safe to remove the groups and place them in the top-level wires section.
\end{enumerate}
\Cref{fig:compilation:remove-holes} shows the resulting program that contains
no groups, interface signals, or control statements.

\paragraph{Code generation}
Each component now contains a flat list of guarded assignments.
The \pass{Lower} pass generates SystemVerilog programs by mapping each
component to a module, generating wires for all the ports, and threading a
clock signal through the design.

\subsection{Compiling Control Statements}
\label{sec:compile:control}

The \pass{CompileControl} pass performs a bottom-up traversal of the control
program, encodes the control flow of each control statement using structural
components, and replaces its use with corresponding compilation group.
This example illustrates the timeline of bottom-up elimination of control statements:

\vspace{1.3ex}
\noindent
\begin{minipage}{0.38\columnwidth}
\begin{lstlisting}
control { par {
  seq { one; two; }
  seq { foo; bar; }
}}
\end{lstlisting}
\end{minipage}%
\hfill
\begin{minipage}{0.3\columnwidth}
\begin{lstlisting}[escapeinside={<@}{@>}]
control { par {
  seq0;
  seq1;
}}
\end{lstlisting}
\end{minipage}
\hfill
\begin{minipage}{0.3\columnwidth}
\begin{lstlisting}[escapeinside={<@}{@>}]
control {
  par0;
}
\end{lstlisting}
\end{minipage}

\noindent
We sketch the \pass{CompileControl} pass's strategies for implementing each control statement in \sys{}.

\paragraph{par}
A \code{par} control block enables all groups inside it and finishes executing
when all groups have signaled \code{done} once.
Since groups may finish executing at different times, the pass generates a
$1$-bit register to save each child group's done signal.
The go signal for each child group is set to high when the value in this
register is $0$.
The done signal for the compilation group is $1$ when all the $1$-bit registers
output $1$.

\paragraph{if}
\sys{}'s semantics dictate that an \code{if} statement executes a group
\code{cond} before reading the value from a port and deciding which branch to
execute.
\code{cond} is supposed to update the value on the port.
The pass generates two $1$-bit registers:
\code|cc| which tracks if \code|cond| has been executed, and
\code|cs| to store the value of the port generated after executing \code|cond|
to ensure that the value of the port is available through the execution of
the branches.
The compilation group enables either branch using the value in \code|cs| and
finishes executing when the branch's done signal is high.

\paragraph{while}
The loop compilation strategy resembles the one for \code{if}.
The group runs the condition group, saves the value from the
condition port to a register, and uses it to either enable the group in
the body.
The compilation group finishes executing when the value of the conditional port
is $0$.

\paragraph{Resetting compilation groups}
Compilation groups reset their internal state to operate correctly within
loops.
The pass generates assignments that reset the value of internal state elements
when a compilation group sets its done signal high.

\subsection{Latency-Sensitive Compilation}
\label{sec:compile:static}

The default compilation pass, \pass{CompileControl}, generates
latency-insensitive finite-state machines (FSMs) when realizing a component's
schedule.
Such latency-insensitive designs allow the execution schedule to uniformly
reason about multi-cycle components and groups.
The cost of this approach, however, is the additional hardware and additional execution
cycles required to coordinate with the interface signals.
Frontend compilers can often provide latency information that
the compiler can exploit to
build smaller and faster hardware.

We implemented a pass that can opportunistically generate latency-sensitive
FSMs when latency information is available.
This pass is best-effort: it only attempts to generate such FSMs when latency
information is available and gracefully falls back to \pass{CompileControl}.
The encapsulation property of groups enables these kinds of best-effort
passes---the compilation pipeline does not have to reason about what is inside
a group to compile it.

The key benefits to this approach are:
(1) frontends can quickly build a functioning end-to-end flow and incrementally
add latency information to generated programs, and
(2) latency-sensitive compilation is \emph{just} an optimization---it can be
disabled, debugged, and interacted with separately from the compilation
pipeline.
To the best of our knowledge, \sys{}'s ability to fluidly mix latency-sensitive
and latency-insensitive compilation is unique.
Prior systems intertwine latency information through the compilation process,
so either everything is statically timed~\cite{vivadohls2017} or nothing is~\cite{dynamic-schedule-hls}.

\Cref{sec:dahlia:static} shows how a frontend can generate latency
information, \cref{sec:eval:dahlia} demonstrates that the pass speeds up
designs by \staticTiming without an area penalty, and \cref{sec:opt:infer-latency} demonstrates how latency information can be automatically inferred
in certain cases.

\paragraph{Compiling seq}
The latency-sensitive compilation pass, \pass{Sensitive}, traverses the control
program bottom-up and opportunistically compiles control statements when all of
the nested groups specify their latency using the \code{static} attribute
(\cref{sec:lang:attributes}):
\\
\begin{minipage}{1.0\linewidth}
\begin{lstlisting}
group one<"static"=1> { ... }
group two<"static"=2> { ... }
control { seq { one; two } }
\end{lstlisting}
\end{minipage}
\\
It generates an FSM with a self-incrementing counter and
enables each group for the specified number of cycles,
and ignores the \code{done} signal from the groups:
\begin{lstlisting}
group static_seq0<"static"=3> {
  one[go] = fsm.out >= 0 && fsm.out < 1 ? 1;
  two[go] = fsm.out >= 1 && fsm.out < 3 ? 1;
  static_seq0[done] = fsm.out == 3 ? 1;
  // Increment the FSM.
  fsm.in = fsm < 3 ? fsm.out + 1;
  static_seq0[done] = fsm.out == 3;
}
\end{lstlisting}
When compiling \code{seq}, \code{par}, or \code{if}  statements, the pass uses
the latency information of the contained groups to generate a \code{static}
attribute for generated compilation group.

The pass demonstrates how \sys{} enables development of small, modular passes that interact with the broader infrastructure.
It is feasible because the IL has a well-defined semantics that lets passes reason independently about the preservation of program semantics.

\section{Optimizing \sys{} Programs}
\label{sec:opt}

We describe the design and implementation of three optimizations
that demonstrate \sys{}'s ability to support control-flow-sensitive
optimizations.

\begin{figure}
  \begin{subfigure}[b]{0.55\linewidth}
    \centering
    \begin{lstlisting}
group let_r0 { r0.in = 0 }
group let_r1 { r1.in = 0 }
group incr_r0 {
  a0.l = r0.out; a0.r = 1;
  r0.in = a0.out; }
group incr_r1 {
  a1.l = r1.out; a1.r = 1;
  r1.in = a1.out; }
    \end{lstlisting}
    \caption{Defined groups. \code|r0| and \code|r1| are registers; \code|a0| and \code|a1| are adders.}
    \label{fig:opt:groups}
  \end{subfigure}
  \hfill
  \begin{subfigure}[b]{0.42\linewidth}
    \centering
    \begin{lstlisting}
seq {
  par {
    let_r0;
    let_r1;
  }
  incr_r0;
  incr_r1;
}
    \end{lstlisting}
    \caption{Schedule with resource sharing opportunities.}
    \label{fig:opt:resource-sharing}
  \end{subfigure}

  \caption{Resource sharing example. Since \code|incr_r0| and \code|incr_r1| do
not run in parallel, they can share their adders.}
  \label{fig:opt}
\end{figure}

\subsection{Resource Sharing}
\label{sec:opt:resource-sharing}

Resource sharing is an optimization that reuses existing circuits to perform
temporally disjoint computations.
For example, if an accelerator needs to perform two add operations that are
never executed in parallel, it can map them to the same physical adder.
\sys{} is uniquely suited to implement such optimizations
which require both control flow facts (if two computations run in
parallel) and structural facts (which physical adder performs an add).

\sys{} implements a group-level resource sharing optimization: if two groups
are guaranteed to never execute in parallel, they can share components.
This pass does not attempt to share stateful components because state is
visible across groups.
Frontends use the \code|"share"| attribute (\cref{sec:lang:attributes}) to
denote that a component is safe to share.
\begin{lstlisting}
component adder<"share"=1> { ... }
\end{lstlisting}
The pass uses the execution schedule of a component to calculate which groups may run in
parallel and uses the encapsulation property of groups to
implement sharing.
It proceeds in three steps:

\paragraph{Building a conflict graph}
A conflict graph summarizes potential conflicts---nodes denote groups and edges
denote that the groups \emph{may} run in parallel.
The pass traverses the control program and adds edges between all children
of a \code{par} block.
For example, in \cref{fig:opt:resource-sharing}, the groups \code|let_r0| and
\code|let_r1| conflict with each other while \code|incr_r0| and \code|incr_r1|
do not.
If the children of the \code|par| block are themselves control programs, the
pass adds edges between the groups contained within each child.

\paragraph{Greedy coloring}
The pass performs a greedy
coloring over the conflict graph to allocate shareable components to each
group.
If two groups have an edge between them, they cannot have the same components.
The result of this step is a mapping from the names of old components to
new components.
For example,
in \cref{fig:opt:groups}, \code|incr_r1| gets the mapping: $a1 \mapsto a0$.

\paragraph{Group rewriting}
In the final step, the pass applies local rewrites to groups based on the
mapping.
The simplicity of this step comes from the encapsulation property of groups---a
rewriter does not have to reason about uses of a component outside the group.

Resource sharing demonstrates \sys{}'s flexibility in analysis and
transformation---passes can recover control flow information from the schedule
and use groups to perform local reasoning.

\subsection{Register Sharing via Live-Range Analysis}
\label{sec:opt:live-range}

Group-local reasoning is insufficient for sharing stateful elements such as
registers; writes to a register in one group are visible in other groups.
To enable register sharing, we implement a
\emph{live-range analysis} that, for each register, determines the
last group in the execution schedule to read from it.
Since the register is guaranteed to never be used afterwards, subsequent groups
can reuse the register.
Live-range analysis is common in software
compilers but is infeasible in RTL languages since the control flow of the
program is not explicit.

The live-range analysis has to contend with two problems:
(1) coping with the \code|par| blocks in the control program, and
(2) inferring which groups read and write to registers.

\paragraph{Parallel control flow graphs}
\begin{figure}
  \begin{subfigure}[b]{0.42\linewidth}
\begin{lstlisting}
seq {
  A;
  if cond.out with G {
    B;
  } else {
    par {
      seq { x0; x1; }
      seq { y0; y1; }}}
  C;
}
\end{lstlisting}
    \caption{\sys{} program.}
    \label{fig:opt:pcfg-code}
  \end{subfigure}
\hfill
  \begin{subfigure}[b]{0.57\linewidth}
    \includegraphics[width=\linewidth]{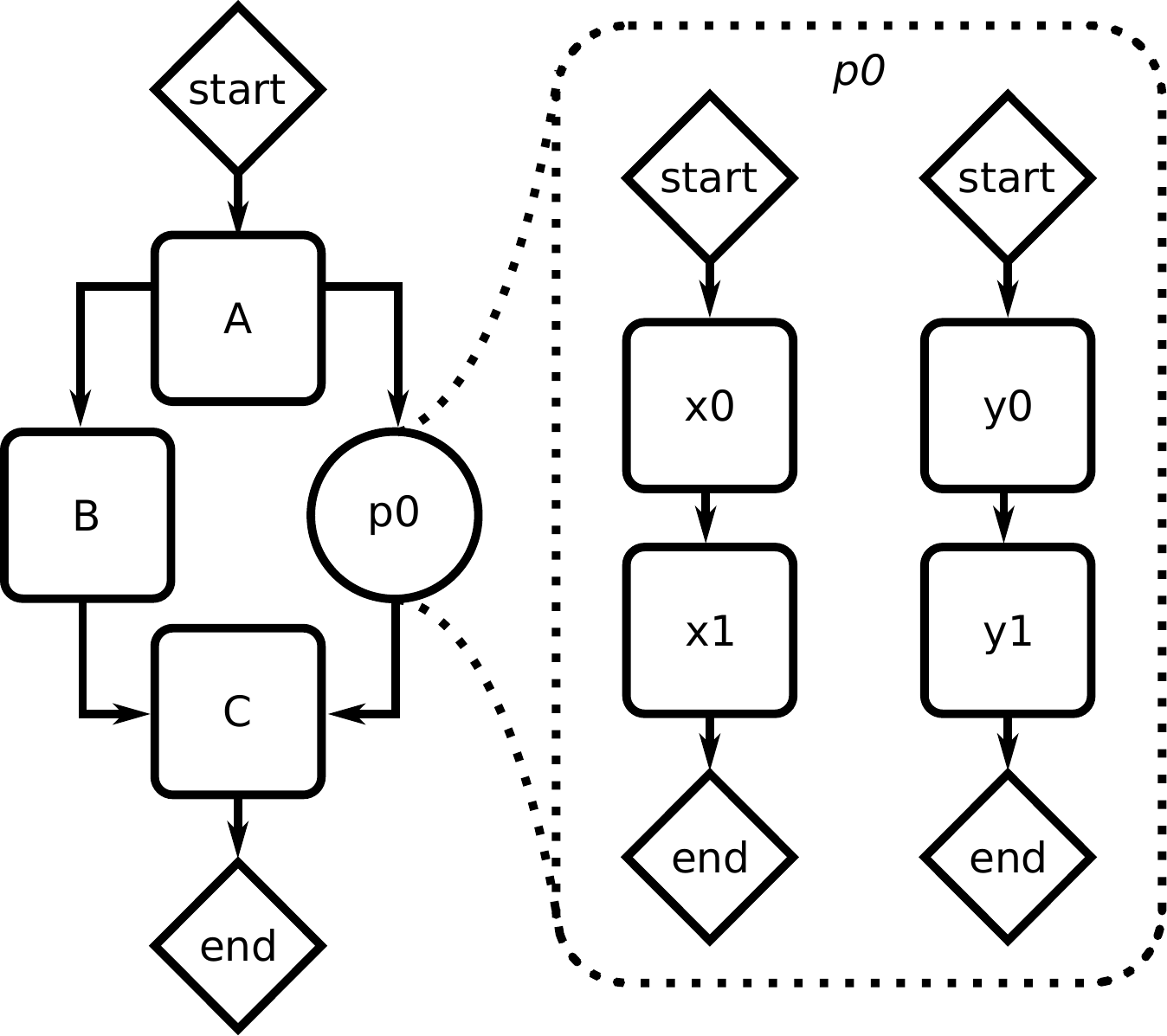}
    \caption{A visual representation of a pCFG.}
    \label{fig:opt:pcfg-diagram}
  \end{subfigure}
  \caption{A \sys{} program along with the corresponding parallel control flow graph (pCFG).}
  \label{fig:opt:pcfg}
\end{figure}

We handle \code|par| blocks using parallel control flow graphs (pCFGs) based
on the work of \citet{explicit-parallelism}. Most control operators in \sys{} map directly
to a traditional CFG. However, \code|par| statements need special handling
since, unlike an \code|if| statement which executes one of its two branches, a
\code|par| statement executes \emph{all} its children.
While writes to a register in a conditional branch \emph{may} be visible after
the \code|if| statement, writes within children of \code|par| blocks are
\emph{always} visible after the \code|par| block.

Parallel CFGs introduce a new kind of node---called a \emph{p-node}---to handle
\code|par| blocks ($p0$ in \cref{fig:opt:pcfg-diagram}).
A p-node represents an entire \code{par} block and recursively contains a set
of pCFGs representing its children.
In \cref{fig:opt:pcfg-diagram} the p-node has two children.

\paragraph{Calculating read and write sets}
\sys{} implements a conservative analysis pass to determine the registers that
groups and p-nodes read from and write to.
Both groups and p-nodes can, in general, contain complex logic, so the pass must
conservatively over-approximate these sets.
The read set is the set of registers a group or p-node \emph{may} read from and
the write set is the set of registers they \emph{must} write to.
The data-flow analysis uses this information to determine the range each
register is alive.

\paragraph{Computing liveness}
The pass uses a standard data-flow formulation to compute the live ranges.
The only
aspect that needs special handling is the children of p-nodes. For these,
we set the live sets at the end of each child to be the set of live registers
coming out of the p-node.

\paragraph{Sharing registers}
The pass uses the liveness information to build a conflict graph where nodes
are registers and edges denote overlapping live ranges.
The pass performs greedy coloring over this graph using registers as colors
and rewrites groups in a similar manner to resource sharing.

\subsection{Inferring Latencies}
\label{sec:opt:infer-latency}

The final optimization pass in the \sys{} compiler attempts to conservatively
infer the latencies of groups and components.
This enables the downstream \pass{Sensitive} pass (\cref{sec:compile:static})
to lower \sys{} programs using more efficient, latency-sensitive finite state
machines.
Consider the following group:
\begin{lstlisting}
component foo<"static"=1> { ... }
group incr {
  f.in = add.out; // f is an instance of foo.
  f.go = 1'd1;
  incr[done] = f.done;
}
\end{lstlisting}
The \sys{} program specifies that the latency of the \code|foo| component is
$1$ using the \code|"static"| attribute.
Given this information, this pass infers that latency of \code|incr| to be
$1$ as well.
It follows a simple rule: if a group's done signal is equal to a components
go signal, and if the component's go signal is set to $1$ within the group,
the latency of the group is inferred to the same as the component.
Such uses of components occur in groups that simply activate one component and
end their execution.

This pass is conservative and only works for simple groups.
Given \sys{}'s design principle---that most of the time frontends generate
simple groups---such passes can be extremely powerful.
Furthermore, such passes can be incrementally improved by adding new rules that
enables the pass to infer latencies for more groups and transparently speed
up programs.
\Cref{sec:eval:systolic} shows that this pass transparently improves the
performance of frontend code.

\section{Case Studies}
\label{sec:study}

We built two compilers that target \sys{} for our case studies.
The first generates systolic arrays~\citep{kung:systolic} for linear algebra
computations.
The second compiles Dahlia~\cite{dahlia}, an imperative programming language
that uses a substructural type system to enable predictable hardware design.
Our goal in both case studies is to demonstrate how \sys{} makes it possible to
quickly bring up good compiler implementations for specialized languages.
We do not aim to beat existing commercial HLS compilers which represent decades
of engineering effort.

\subsection{Systolic Array Generator}
\label{sec:study:systolic}

\begin{figure}
    \begin{center}
    \includegraphics[width=0.85\linewidth]{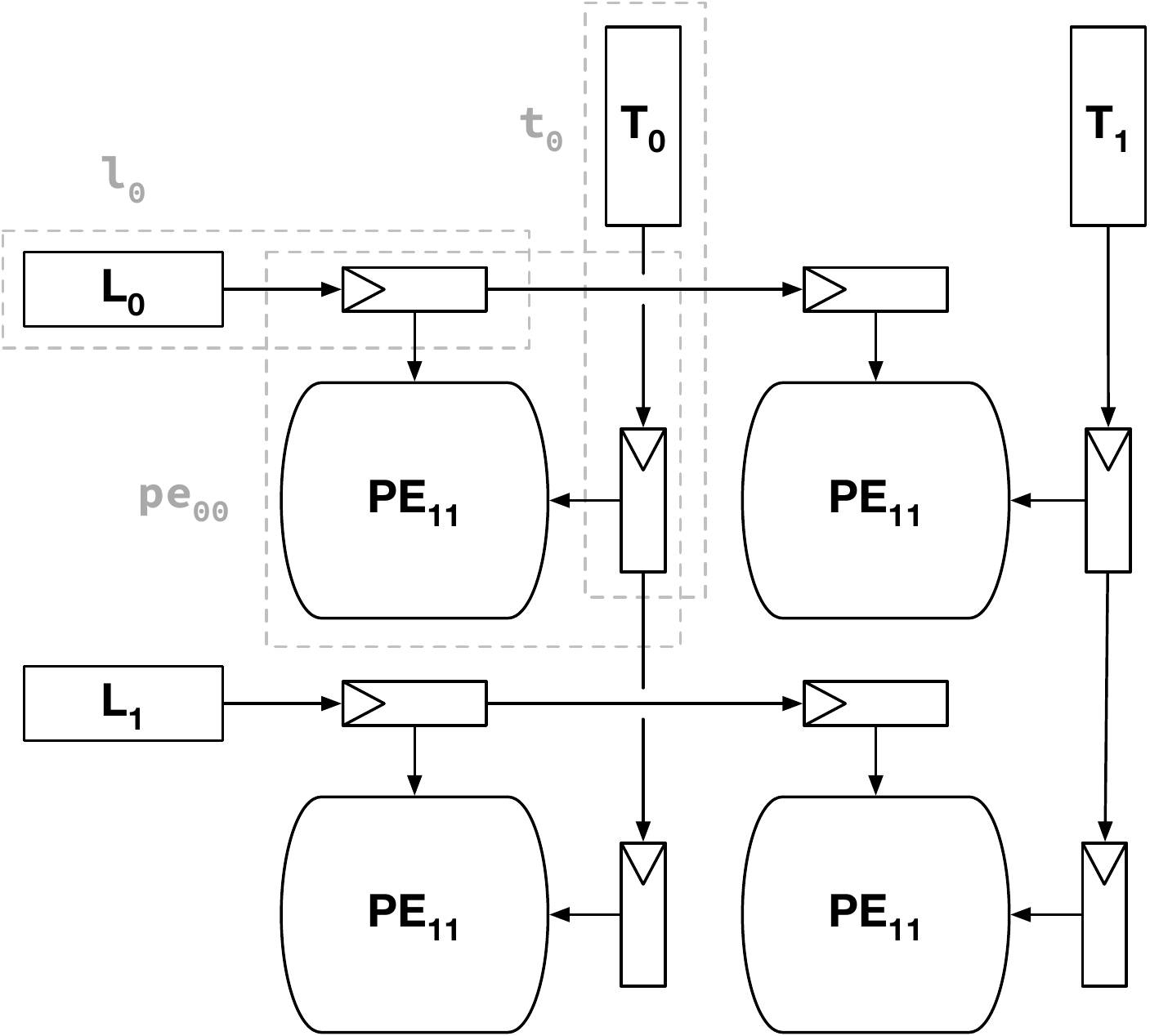}
    \end{center}
    \caption{Architecture for a 2$\times$2 by 2$\times$2 matrix-multiply systolic array. Highlighted boxes show some of the groups used by the control.}
    \label{fig:systolic-arch}
\end{figure}

Systolic arrays~\cite{kung:systolic} are a class of architectures that exploit data reuse.
They power
the recent wave of state-of-the-art linear algebra accelerators for machine
learning~\cite{tpu, brainwave}.
\Cref{fig:systolic-arch} shows an example systolic array.
In every time step, data moves from left-to-right and top-to-bottom, while the
processing elements (PEs) in the grid perform computations on the data streams.
Systolic arrays can maintain a high throughput because data is reused between
PEs.

However, generating a custom systolic array implementation is challenging:
producing RTL directly requires generating complex custom control hardware, and
systolic arrays' unique parallelism pattern can be challenging to express
in HLS \cxx~\cite{polysa, susy}.
We implement a systolic array generator using \sys{} in only
\systolicLOC{} LOC of Python and approximately $40$ person-hours of effort.
The generator can produce arrays with arbitrary dimensions and arbitrary PEs
which are implemented as \sys{} components themselves.

\paragraph{Input}
The systolic array generator takes the dimensions of the matrix block
and a \sys{} component that implements the PE.
For a matrix multiply accelerator, for example, the PE consists of a
multiply--accumulate (MAC) unit.
It generates a systolic array that matches the dimensions of matrix block.

\paragraph{Architecture}
\Cref{fig:systolic-arch} shows the desired architecture for a 2$\times$2
systolic array.
The design consists of several groups highlighted in the figure.
The groups that surround a PE implement \emph{data movement:} the groups on
the edges move the data from the input memories to registers,
and the ones in the middle move the data along the fabric.
Finally, the compute groups perform the computation in the PE and write their
results to an internal register.

\paragraph{Generating \sys{}}

\begin{figure}
\begin{lstlisting}
seq {
  par { t0; l0; } // Move data from memories
  par { pe_00; }  // Run the first PE
  // Move data from memories and from registers
  par { t0; t1; l0; l1; pe_00_down; pe_00_right; }
  // Execute first PE and PEs on diagonals
  par { pe_00; pe_01; pe_10; }
  // Next step...
  par { t1; l1; down_00; down_01; right_00; right_10; }
  par { pe_01; pe_10; pe_11; }
  par { down_01; right_10; } par { pe_11; }
}
\end{lstlisting}
\caption{Control generated for 2x2$\times$2x2 matrix-multiply.
Execution interleaves data movement and PE execution.}
  \label{lst:systolic-control}
\end{figure}

To target \sys{}, the systolic array generator needs to (1)
instantiate PEs, (2) create the relevant groups, and (3) define the
control for the systolic array.
The compiler performs (1) and (2) using templates.
For each PE, the compiler also instantiates the surrounding
input registers and connects them to registers in the previous PE.
Finally, it defines groups to move the data and perform the computation.

The next step is generating the control.
\Cref{lst:systolic-control} shows the control statements generated for a
2$\times$2 systolic array.
At each time step, the compiler enables all the data movement
groups to move the data to input registers of each PE, and then in sequence,
enable all the PEs with valid inputs.
This schedule implements the classic systolic data flow that implements matrix
multiplication which shifts the input data by one step in each dimension.
The generated control accounts for invalid data and selectively enables
data movement and compute groups when the input data streams start and end.

\paragraph{Inferring latencies}
The systolic array generator does not generate any \code|"static"| annotations.
However, the \sys{} compiler is able to completely infer the latency
(\cref{sec:opt:infer-latency}) of a generated systolic array when the
processing element provides its latency.
This means that the generator, by virtue of using the \sys{} compiler,
automatically supports both latency-sensitive and latency-insensitive systolic
arrays.

\paragraph{Debugging with \sys{}}
In an initial version, the generator prematurely enabled data
movement groups causing the systolic array to compute the wrong result.
While debugging the kernel, it was easy to spot this mistake in the control
program.
This demonstrates a key quality-of-life improvement when using the \sys{}
infrastructure to build accelerator generators---control logic bugs can be
caught by investigating the execution schedule.

\subsection{The Dahlia Compiler}
\label{sec:study:dahlia}

Dahlia~\citep{dahlia} is a recently proposed general-purpose language for designing accelerators that resembles traditional C-based HLS.
It differs from traditional HLS by adding a substructural type system that constrains the language to rule out programs that lead to inefficient hardware.
The original Dahlia compiler generates \cxx with annotations for the commercial
Vivado~HLS~\cite{vivadohls2017} toolchain.

In this case study, we build a new compiler for the Dahlia language that
generates hardware using \sys{}, eliminating the dependence on a monolithic,
closed-source HLS backend and allowing greater control over the generated
architecture.
The goal is not to outperform the Vivado~HLS backend; instead, we aim to show that \sys{}
makes it possible to exploit Dahlia's unique semantics to build a compiler that
is far simpler than a full-fledged C-to-RTL toolchain.

\paragraph{Lowered Dahlia}

Dahlia is a simple imperative language extended with high-level convenience
features such as memory partitioning, loop unrolling, and logical array
indexing.
We elide the details of the first step of compilation that unrolls loops and
compiles accesses to partitioned memories.
We refer interested readers to \href{https://github.com/cucapra/dahlia/tree/3acddf5277beba750065564f9c9206b55d58ae18/src/main/scala/passes}{our implementation}.

Our explanation focuses on compiling Dahlia programs that use a small
set of constructs: variables, unpartitioned memories, \code{while} loops,
conditionals, and Dahlia's two novel composition operators:
\emph{unordered} composition (\code{;}) and \emph{ordered} composition
(\code{---}).

In Dahlia, memories and variables have an associated type and can be updated
with assignment syntax:
\begin{lstlisting}[language=fuse]
let x: ubit<32> = 1; x := 2;
let arr: ubit<32>[10]; arr[1] := 3;
\end{lstlisting}
Dahlia's \emph{unordered} composition operator allows backends to parallelize
computations while preserving data flow:
\begin{lstlisting}
x = 1; y = 2  // can occur in parallel
\end{lstlisting}
In contrast, Dahlia's \emph{ordered} composition operator requires backend to
execute statements in a sequence:
\\
\begin{minipage}{1.0\linewidth}
\begin{lstlisting}[language=fuse]
x = 1
---
x = 2
\end{lstlisting}
\end{minipage}
\\
Ordered composition does not reason about explicit clock cycles.
Instead, it imposes a partial order over the execution of program statements by
reasoning about \emph{logical timesteps}.
Lowered Dahlia also supports standard imperative \code{while} loops and
\code{if} conditionals.

\paragraph{Generating \sys{}}
The \sys{} backend for Dahlia is a bottom-up pass that compiles each
expression by instantiating groups and scheduling them using the control
language.

For example, for this Dahlia program:
\begin{lstlisting}[language=fuse]
let x = 0
---
if (x > 10) { x = 1 } else { x = 2 }
\end{lstlisting}
The \sys{} backend generates a group for each statement:
\begin{lstlisting}
group init_x { x.in = 0; init_x[done] = x.done; }
group one { x.in = 1; one[done] = x.done; }
group two { x.in = 2; one[done] = x.done; }
group cond { gt.left = x.out; gt.right = 10; cond[done] = 1; }
\end{lstlisting}
And schedules them using the following control program:
\begin{lstlisting}
seq {
  init_x;
  if gt.out with cond { one } else { two }
}
\end{lstlisting}
The \sys{} backend has a one-to-one mapping for the language constructs in
lowered Dahlia and the \sys{} control language:
memory and variable assignments generate groups representing the update,
ordered composition becomes \code{seq},
unordered composition becomes \code{par},
loops and conditionals map to \code{while} and \code{if}.

\paragraph{Interfacing with black-box RTL}
Dahlia's HLS backend uses a vendor-provided header to implement custom
math functions such as square root.
The HLS compiler connects definitions within such headers to black-box
RTL code.
In order to interact with black-box RTL components, \sys{} programs can
provide external definitions:
\begin{lstlisting}
extern "sqrt.sv" {
  component sqrt(left: 32, right: 32, go: 1) -> (
    out: 32, done: 1
  );
}
\end{lstlisting}
External definitions do not provide an implementation; instead the \sys{}
compiler links in the corresponding RTL program, in this case \code|sqrt.sv|,
during code generation.
External components can be used like any other component:
\begin{lstlisting}
group foo {
  sqrt0.left = 10; sqrt0.right = 20;
  sqrt0.go = !sqrt0.done ? 1;
  foo[done] = sqrt0.done
}
\end{lstlisting}

\paragraph{Latency annotations}
\label{sec:dahlia:static}
Most operations in a Dahlia program have a precise latency---register writes take one cycle, multiplies take four cycles, etc.
The \sys{} backend uses this information to annotate the latency of each group
with the \code|"static"| attribute. Some operations, such as the RTL primitive to calculate the square-root, take a
data-dependent number of cycles, so groups using them omit latency
information.
Since the \sys{} compiler gracefully handles mixed latency-sensitive and
latency-insensitive groups, we do not need to change anything else.

\subsection{Summary}
In our experience, a \sys{}-based compiler requires three ingredients:
(1) the abstract architecture for the domain,
(2) a mapping from source constructs to \sys{} constructs, and
(3) a strategy to generate \emph{groups} and \emph{control}.
For Dahlia, the architecture corresponded directly with the control
language; for systolic arrays, we used a templated
design with a latency-insensitive interface.
In both compilers, we used groups and control to modularize and compose data
flow graphs, which is not possible when generating RTL directly.

\section{Evaluation}
\label{sec:eval}

We evaluate \sys{} by generating accelerators using the frontends in the
previous section and answering three questions:
\begin{itemize}
\item
Can we build a simple compiler that generates performant specialized architectures?

\item
Can we use \sys{}
to generate reasonable hardware in a general-purpose, HLS-like domain?

\item
What is the effect of control-flow-sensitive optimizations implemented in the
\sys{} compiler?

\end{itemize}
We compare \sys{}-generated accelerators to Vivado HLS, a commercial HLS tool
that represents decades of engineering effort.
Our aim is not to beat HLS at its own game but instead achieve the same
performance regime with much lower effort.

\subsection{Systolic Arrays}
\label{sec:eval:systolic}

\begin{figure}
  \centering
  \begin{subfigure}[b]{\linewidth}
    \centering
    \includegraphics[width=0.9\linewidth]{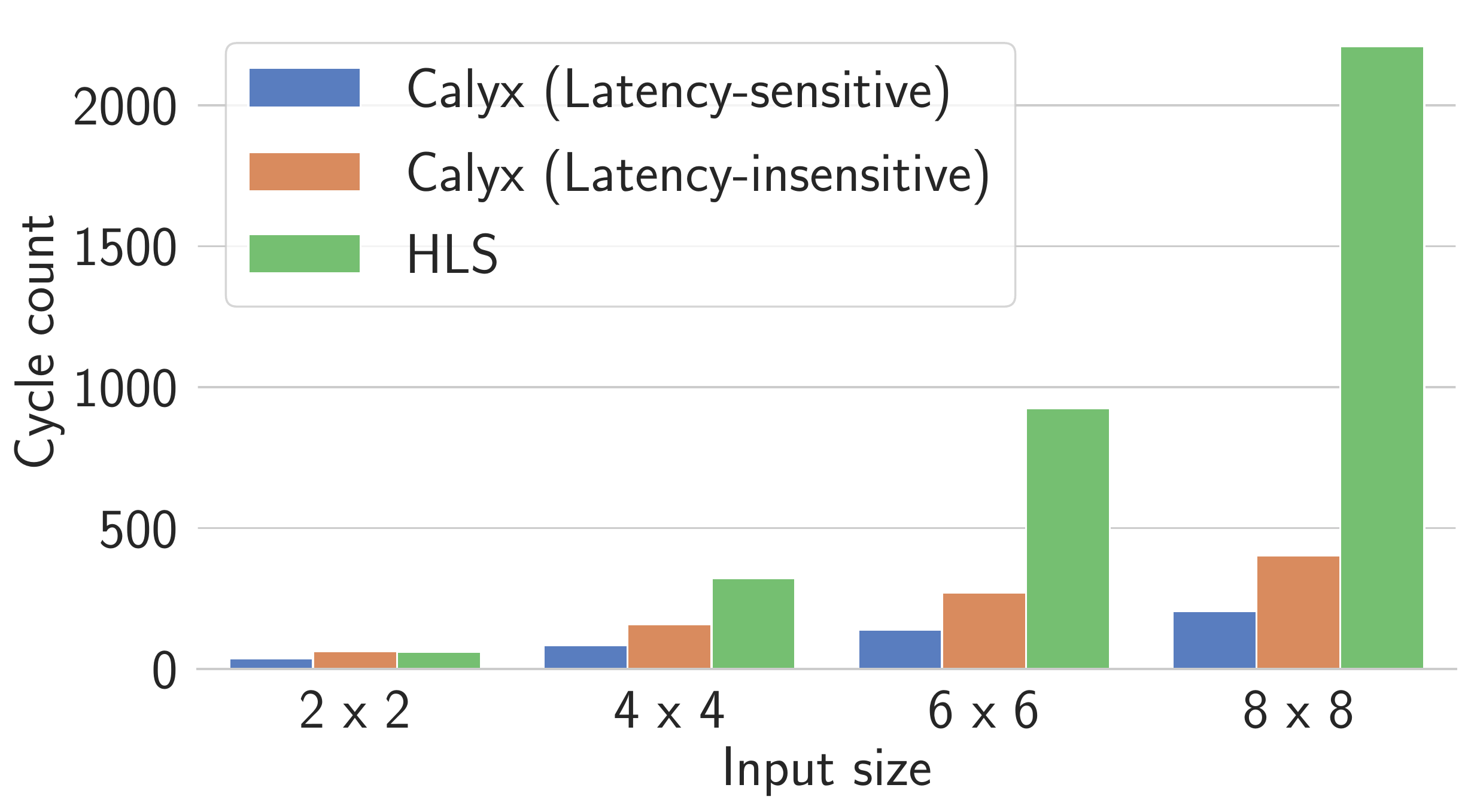}%
    \vspace{-5pt}
    \caption{Absolute cycle counts.}
    \label{fig:systolic-eval:cycles}
    \vspace{6pt}
  \end{subfigure}
  \begin{subfigure}[b]{\linewidth}
    \centering
    \includegraphics[width=0.9\linewidth]{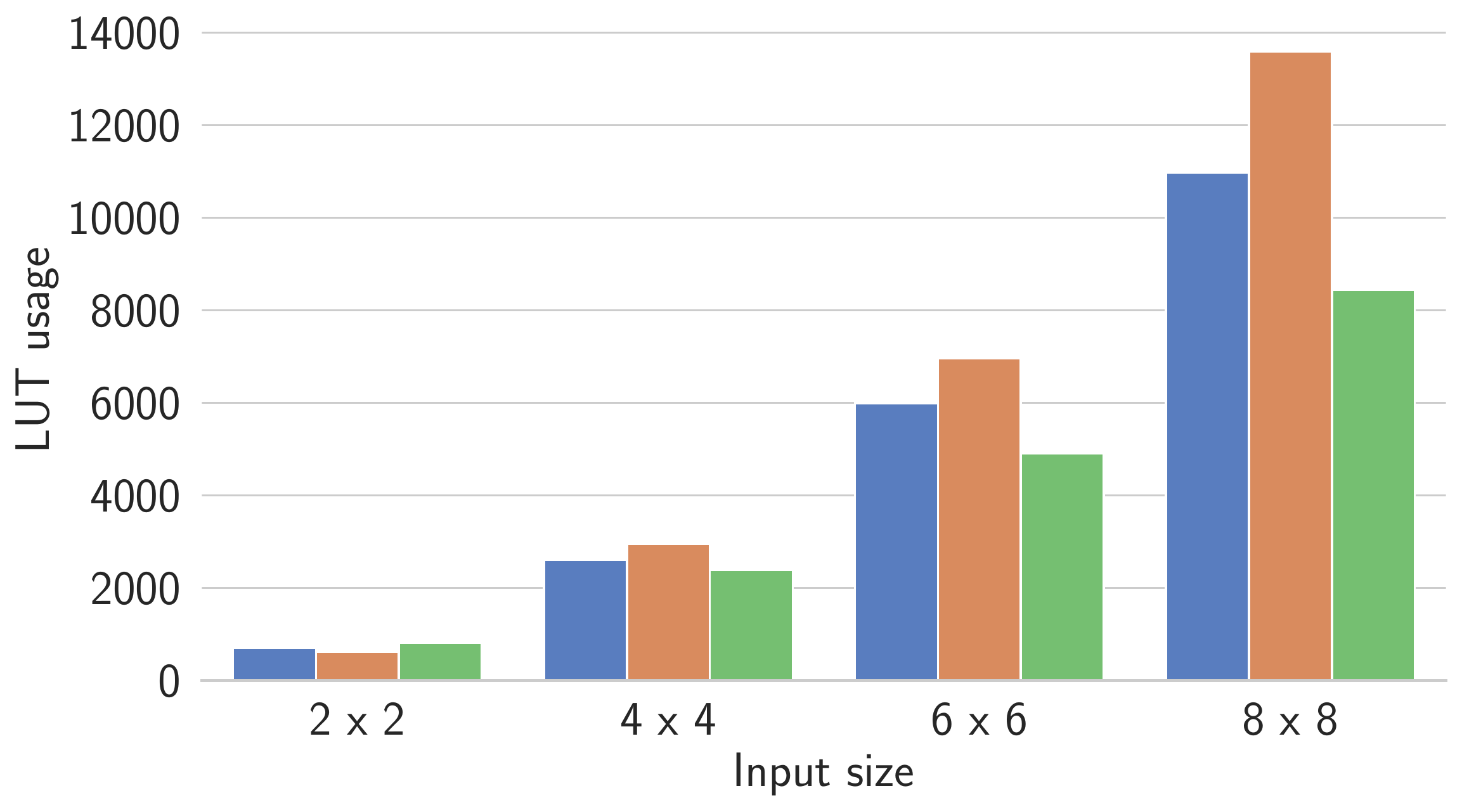}%
    \vspace{-5pt}
    \caption{Absolute LUT usage.}
    \label{fig:systolic-eval:luts}
  \end{subfigure}
  \caption{Resource and cycle count comparison matrix multiply implementation HLS and as systolic arrays.}
  \label{fig:sytolic-eval}
\end{figure}

To the best of our knowledge, Vivado HLS does not automatically
infer systolic arrays from loop nests.
Instead, programmers need to rewrite their program to coerce the
compiler into generating precisely the hardware they want.
\sys{} advocates for a more domain-specific approach---instead of relying on
black-box compilers to infer hardware structures, design new DSLs that
automatically synthesize them.
We study the performance characteristics of the \sys{}-based systolic array
generator (\cref{sec:study:systolic}).

\paragraph{Evaluation setup}
We generate hardware designs for matrix multiplication kernels ranging from
$2\times2$ to $8 \times 8$.
For each configuration, we generate a systolic array using the \sys{}-based
generator and implement a straightforward matrix-multiply kernel in Vivado
HLS that fully unrolls the outer two loops.
For the \sys{} designs,
we collect the number of cycles by simulating the design in
Verilator~\cite{verilator} (v4.108) and get resource estimates by synthesizing
designs with Vivado~\cite{vivadohls2017}, targeting
\fpgaSeries~\fpgaPartname~FPGA at a $7$ns clock period.
For the HLS designs, we report the latency and resource estimates from the
HLS report.
We compare the cycle counts (\cref{fig:systolic-eval:cycles}) and the LUT
usage (\cref{fig:systolic-eval:luts}) of the designs.
We report the characteristics of systolic arrays compiled with
the \pass{Sensitive} pass (Latency-sensitive) and those without (Latency-insensitive).

\paragraph{Comparison against HLS}
Compared to HLS-based designs, \sys{}-based systolic arrays are faster by
a geometric mean of \systolicArrayCycles and take \systolicArrayLut more LUTs.
For the largest input size, the systolic array is \systolicArrayLargeCycles
than the HLS implementation while using \systolicArrayLargeLut
LUTs.

\paragraph{Latency-sensitive compilation}
The systolic array generator does not generate any \code|"static"| annotations
used by the \pass{Sensitive} pass.
It instead relies on the \sys{} compiler to infer these attributes
(\cref{sec:opt:infer-latency}).
On average, \pass{Sensitive} makes designs \systolicArraySensitiveCycles faster
and \systolicArraySensitiveLut smaller.

\paragraph{Discussion}
Our systolic array case study demonstrates how a language designer can quickly
experiment with architectural designs that are harder to express in traditional
HLS tools.
Without extensive engineering effort, the specialized approach can outperform a general-purpose HLS compiler.

\subsection{Dahlia}
\label{sec:eval:dahlia}

\begin{figure}
  \centering
  \begin{subfigure}[t]{\linewidth}
    \centering
    \includegraphics[width=\linewidth]{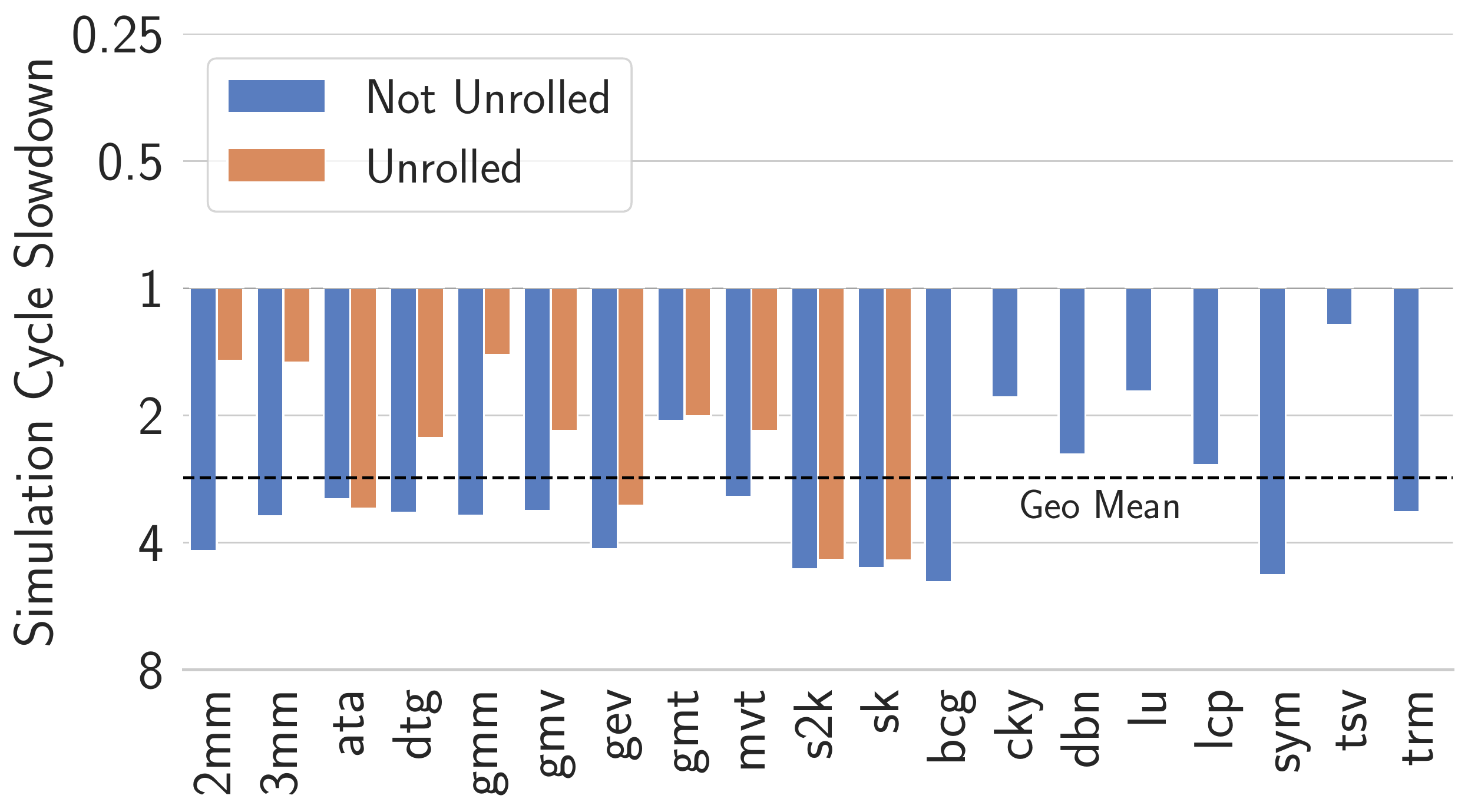}%
    \vspace{-5pt}
    \caption{Cycle slowdown of \sys{} designs compared to Vivado HLS. Designs below the \textit{y}-axis are slower.}
    \label{fig:dahlia-eval:cycles}
    \vspace{6pt}
  \end{subfigure}
  \begin{subfigure}[t]{\linewidth}
    \centering
    \includegraphics[width=\linewidth]{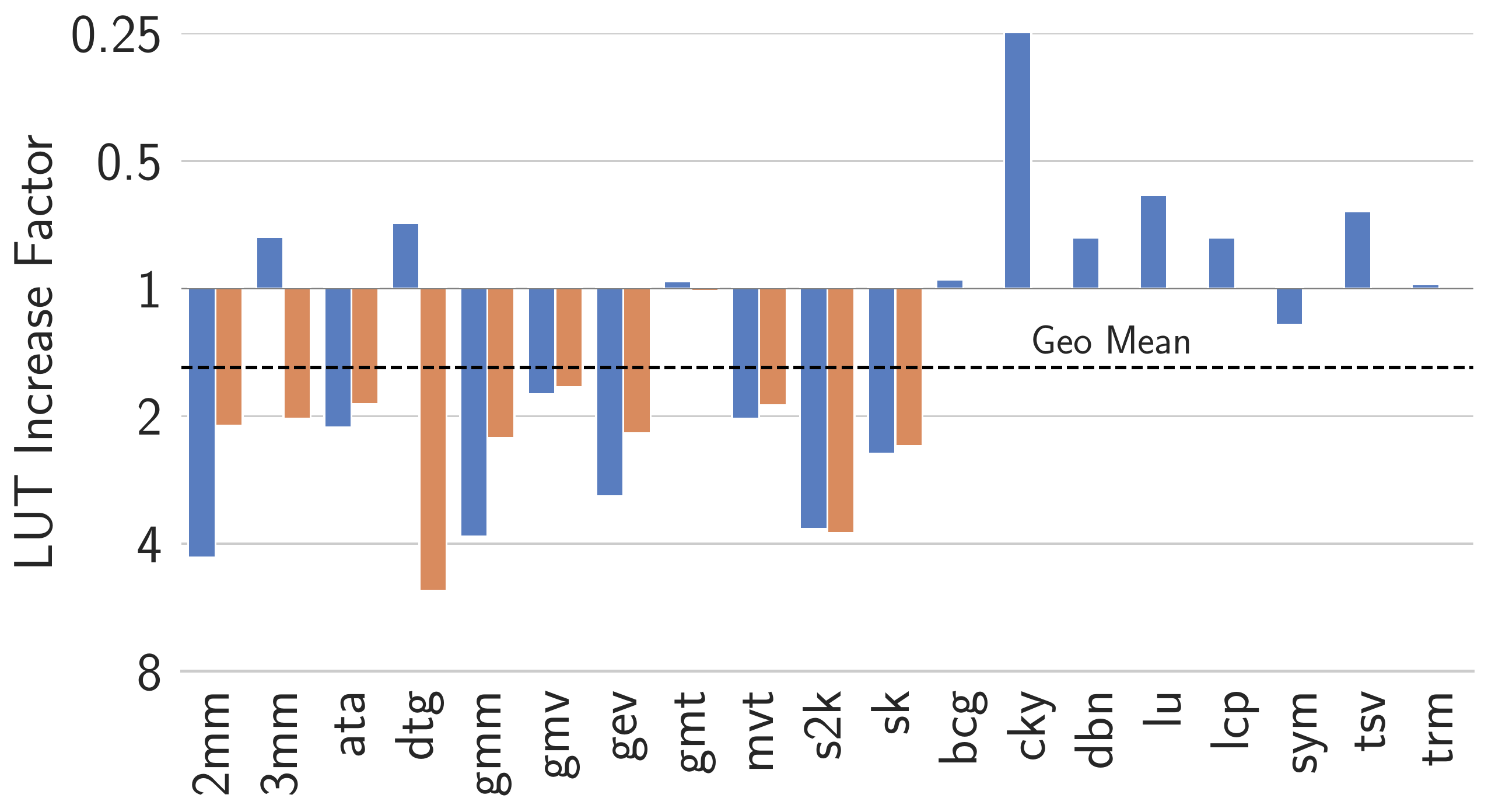}%
    \vspace{-5pt}
    \caption{LUT increase of \sys{} designs over Vivado HLS. Designs below the \textit{y}-axis are larger.}
    \label{fig:dahlia-eval:luts}
  \end{subfigure}

  \caption{%
Resource and cycle count comparison for Dahlia-generated \sys{} designs and HLS
designs for PolyBench benchmarks.
Missing unrolled bars indicate that the benchmark was not unrollable in
Dahlia.}
  \label{fig:dahlia-eval}
\end{figure}

We built the Dahlia-to-\sys{} compiler in 2011 LOC of Scala.
This includes extensions to the Dahlia compiler that add passes to lower Dahlia
specific constructs as well as the backend to generate \sys{} from lowered
Dahlia.

\paragraph{Evaluation setup}

We compare the \sys{}-generated RTL against the original Dahlia
compiler~\cite{dahlia}, which emits annotated \cxx and relies on Vivado~HLS
to generate hardware designs.
We implement all 19 kernels from the linear algebra category of the PolyBench~\cite{polybench} benchmark suite and, for the 11 benchmarks Dahlia's type system allows it, unroll the loops to unlock parallelism.
We use the same setup as in \cref{sec:eval:systolic} to gather numbers.

We also evaluate the effects of the latency-sensitive compilation (\cref{sec:compile:static}).
We run each benchmark with the \pass{Sensitive} pass enabled and disabled,
following the same synthesis and measurement workflow above.

\paragraph{Comparison against HLS}

We collected cycles counts (\cref{fig:dahlia-eval:cycles})
and LUT usage (\cref{fig:dahlia-eval:luts}) for
each benchmark with all optimizations turned on and normalized them to the
corresponding Vivado HLS implementation.
For the unrolled designs, we normalize against the corresponding unrolled
HLS designs.
Since DSP and BRAM usage is almost identical for all benchmarks, we elide them.

On average, the \sys{} generated designs are \polybenchCycles slower than the
designs generated by Vivado~HLS and use \polybenchArea more LUTs. For the
unrolled designs, \sys{} comes closer to HLS execution time, being
\polybenchUnrolledCycles slower while taking \polybenchUnrolledArea more LUTs.
Vivado~HLS is a heavily optimized toolchain that incorporates state-of-the-art
optimizations and is designed to perform well on the kinds of nested loop nests
we evaluated.

\paragraph{Latency-sensitive compilation}
\Cref{fig:dahlia-eval:sensitive} shows the effect of the \pass{Sensitive} pass
(\cref{sec:compile:static}) on the Dahlia-to-\sys{} compiler.
Enabling the optimization reduces execution time on average by \staticTiming
without significantly changing the resource usage.

\paragraph{Discussion}
Despite its simplicity, the Dahlia frontend for \sys{} can already generate designs
 that are within a few factors of the performance of a heavily
optimized, commercial HLS toolchain.
Part of the reason is that Dahlia is a far simpler language than \cxx, which makes a narrowly focused compiler tractable to build.
This is the use case for \sys{}---rapidly designing compilers for specialized
languages and achieving good performance quickly.

We see adding traditional HLS-focused optimizations to \sys{}, such as SDC
scheduling~\cite{sdc-sched}, as the main avenue to close the gap with Vivado~HLS.

\subsection{Effects of Optimization}
\begin{figure}
  \centering
  \begin{subfigure}[b]{\linewidth}
  \includegraphics[width=\linewidth]{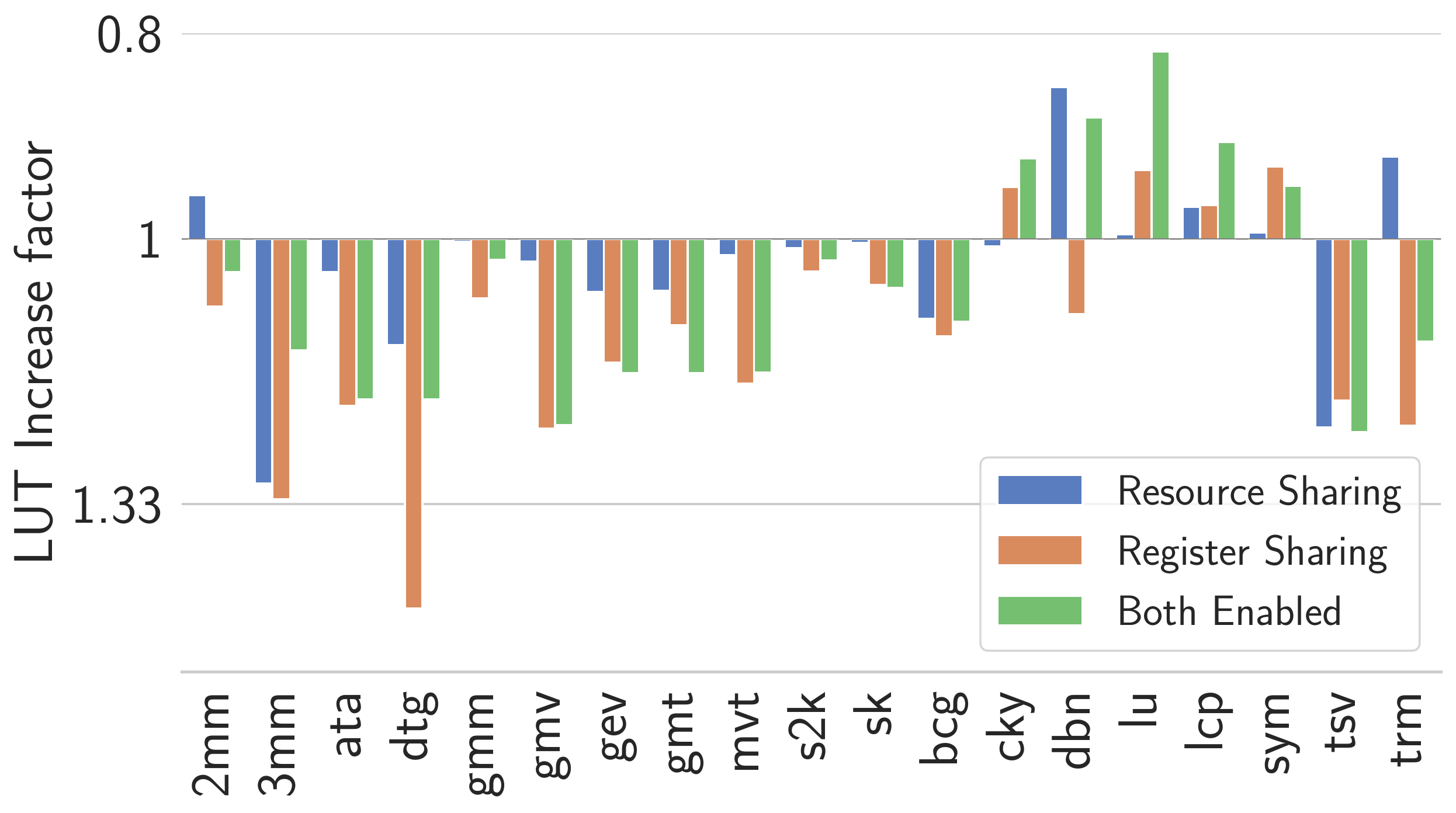}%
  \vspace{-5pt}
  \caption{LUT increase from resource sharing and register sharing.}
  \label{fig:optimizations-eval:lut}
  \vspace{6pt}
  \end{subfigure}
  \begin{subfigure}[b]{\linewidth}
  \includegraphics[width=\linewidth]{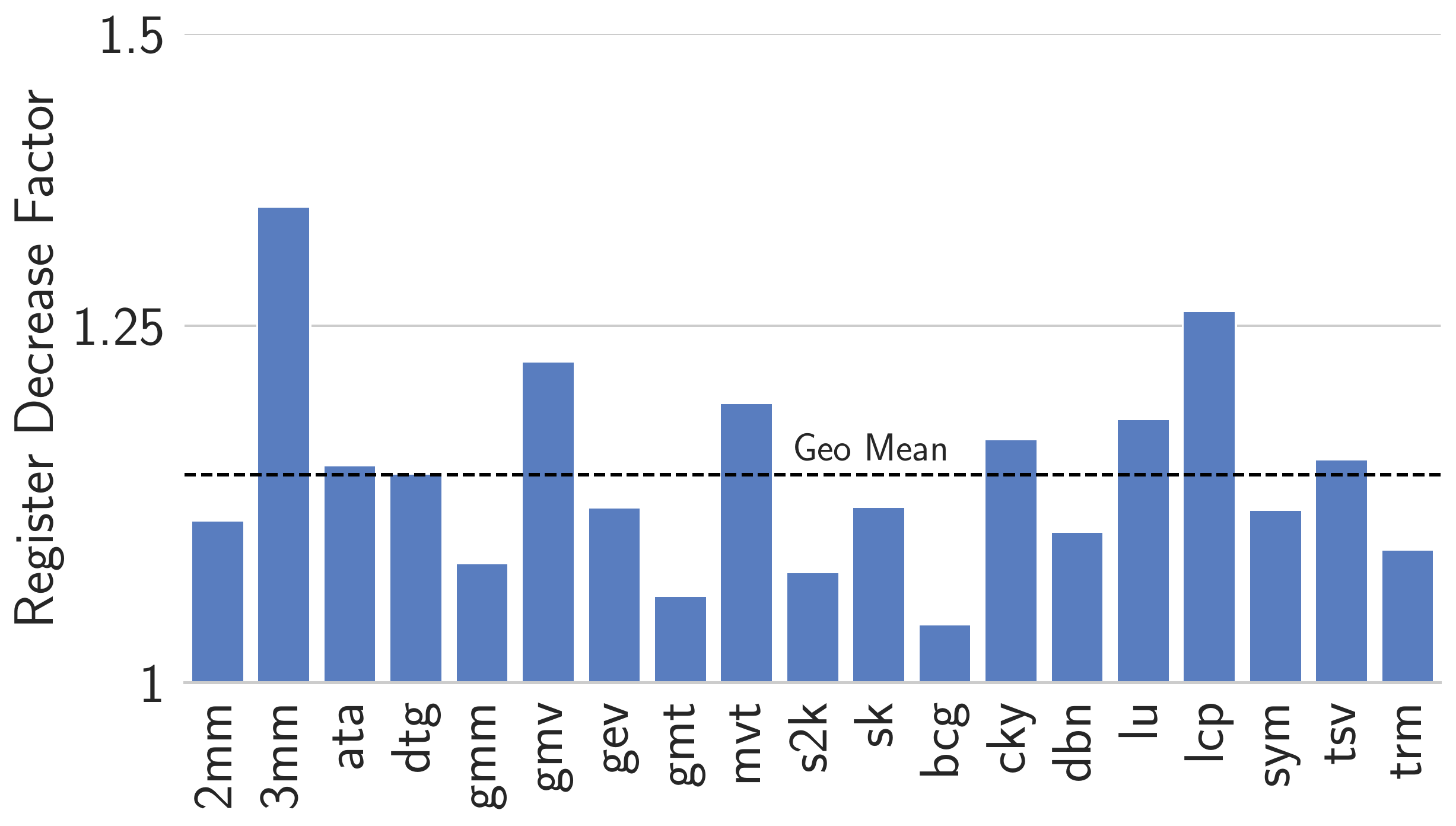}%
  \vspace{-5pt}
  \caption{Register decrease from the register sharing optimization.}
  \label{fig:optimizations-eval:regs}
  \vspace{6pt}
  \end{subfigure}

  \begin{subfigure}[t]{\linewidth}
    \centering
    \includegraphics[width=\linewidth]{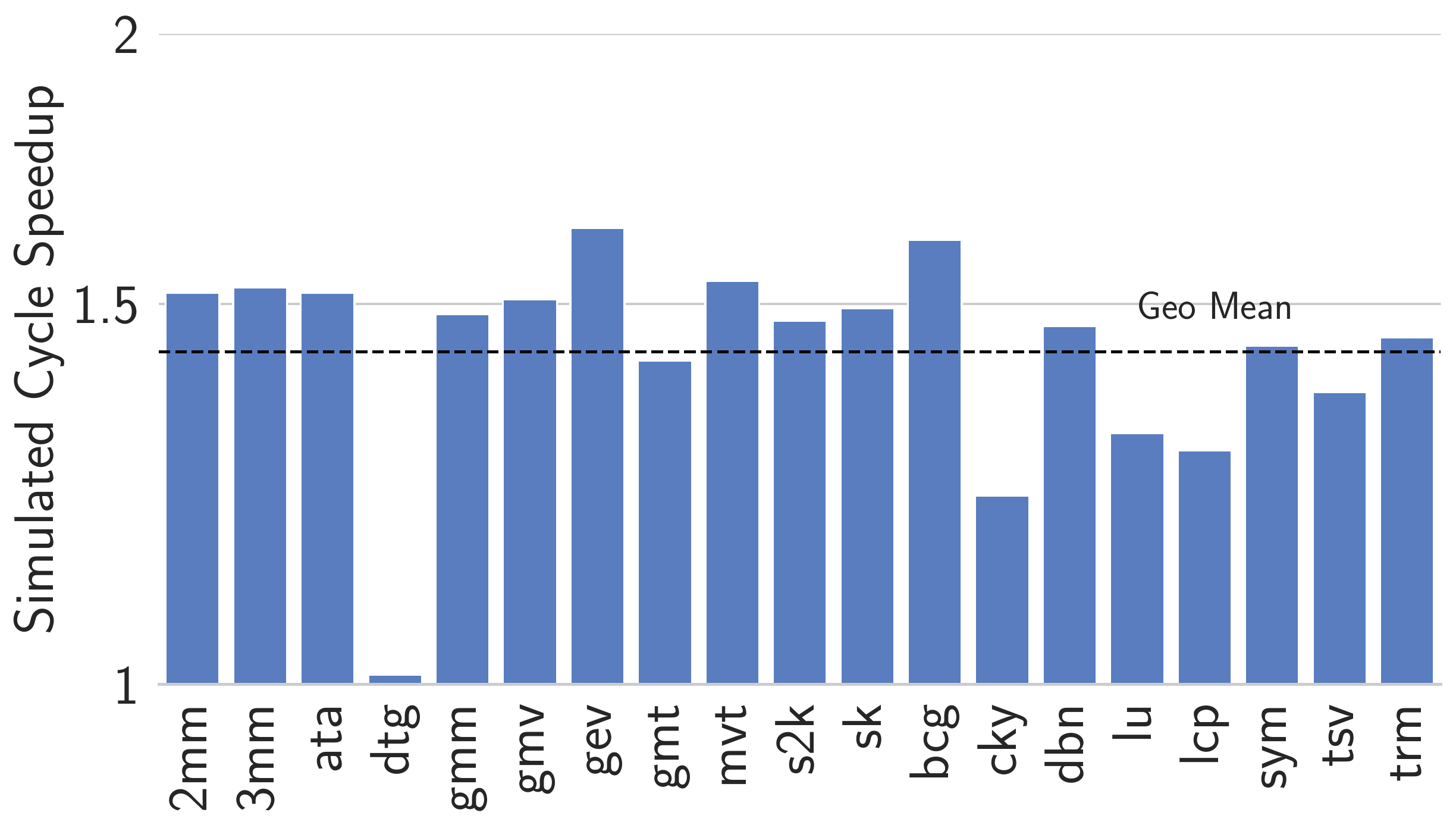}%
    \vspace{-5pt}
    \caption{Speedup from using latency-sensitive compilation.}
    \label{fig:dahlia-eval:sensitive}
  \end{subfigure}

  \caption{Effects of optimization passes. All graphs use logarithmic scales.}
\end{figure}
To demonstrate \sys{}'s ability to express control-flow based optimizations, we
wrote a resource sharing pass (\cref{sec:opt:resource-sharing}) and a register
sharing pass (\cref{sec:opt:live-range}).
We perform an ablation study to characterize their effects on the final
designs.

\Cref{fig:optimizations-eval:lut} reports the resource utilization of
PolyBench benchmarks in three configurations:
(1) resource sharing enabled,
(2) register sharing enabled, and
(3) both resource sharing and register sharing turned on.
We normalize the resource counts against baselines with both passes disabled.

While both optimization passes find opportunities to share hardware
components, there is not a uniform drop in the LUT usage.
On average, the resource sharing pass increases LUT usage by
\optsRsLutGmeanIncrease{} and the register sharing pass increases LUT usage by
\optsMrLutGmeanIncrease{}.
Sharing hardware components causes additional
multiplexers to be instantiated which makes the resource usage worse in some cases.
We plan to implement a heuristic cost model to decide which components are
worth sharing (\cref{sec:future-work}).

\Cref{fig:optimizations-eval:regs} shows the effects of the register sharing
pass on the number of registers used in the designs.
On average, the pass reduces register usage by \optsMrRegGmeanDecrease{} and
finds register sharing opportunities in every benchmark.
Registers, compared to multiplexers, are more expensive in ASIC processes which
represents another opportunity for heuristics in a future
version of the \sys{} compiler.

\subsection{Compilation Statistics}
For the largest PolyBench design (\code|gemver|) \sys{} takes $0.06$ seconds to
generate RTL, compared to $26.1$ seconds for the Vivado HLS compiler.
The largest \sys{} design is the $8\times8$ systolic array which contains
\systolicFutilCells~cells, \systolicFutilGroups~groups, and
\systolicFutilEnables~control statements.
The \sys{} compiler generates \systolicVerilogLOC{} LOC of SystemVerilog
in \systolicFutilCompilerTime{} seconds for this design.

\section{Related Work}

Intermediate representations (IRs) for hardware generation have been a topic of
detailed study.
\sys{} differs from past work because
it is not tied to a specific hardware generation methodology as in IRs for
HLS compilers~\cite{autopilot, legup},
it represents programs at a higher level of abstraction than IRs for RTL
design~\cite{firrtl, coreir},
and it provides precise control over scheduling logic generation unlike
Bluespec~\cite{nikhil:bluespec}.

\paragraph{Bluespec}
Bluespec~\cite{nikhil:bluespec} is an HDL that uses guarded atomic actions to
enable compositional hardware design.
The Bluespec compiler detects conflicts between such actions,
generates a parallel execution schedule,
and dynamically aborts rules on conflicts.
\sys{} requires no implicit dynamic scheduling; it provides explicit control over the
execution schedule using its control language.

\paragraph{Halide}
Halide~\cite{ragan-kelley:halide} is an image processing DSL that pioneered the
separation of algorithmic specifications from the implementation schedule to
facilitate performance tuning, and follow-on work has shown how to compile
Halide-like languages to hardware~\cite{halide-hls,hegarty:darkroom,heterocl}.
Halide schedules represent optimization strategies, such as loop tiling, that
do not affect the algorithm's semantics.
\sys{}'s concept of a schedule is different: it orchestrates and orders the
invocation of hardware components and as such determines the program's
semantics.
\sys{}'s schedules are appropriate for expressing \emph{implementations} of
optimizations like loop tiling performed by high-level DSL compilers.

\paragraph{Software IRs}
Some hardware generators repurpose software IRs such as LLVM~\cite{legup,autopilot,shanghls, llvm},
GCC's internal IR~\cite{bambu},
and SUIF~\cite{suifhls}.
\sys{} is different from these approaches since it does not limit frontend
compilers to sequential, C-like semantics.
It can represent both hardware resources and fine-grained parallelism that these representations lack.

\paragraph{IRs for HLS}
Several HLS compilers include IRs that extend their sequential input languages with representations of parallelism.
\muir{}~\cite{muir} uses a task-parallel representation,
SPARK~\cite{gupta:spark} targets speculation and parallelism optimizations,
CIRRF~\cite{guo:cirrf} provides primitives for pipelining,
and \citet{wu:hierarchical} propose a hierarchical CDFG representation.
\sys{} differs from these IRs by providing lower-level control
primitives to explicitly represent hardware resources and avoiding ties to a traditional HLS setting.

Another category of HLS IRs uses finite state machines (FSMs) to model programs' execution schedules at the cycle level~\cite{ahir, synasm, dutt:intermediate}.
While such FSM representations are reminiscent of \sys{}'s control language,
these IRs impose restrictions on the timing behavior of the operations inside
the FSMs.
\sys{} imposes no such restrictions and can compose arbitrary RTL programs
while providing an interface to generate optimized latency-sensitive designs
when possible.

\paragraph{Languages with hardware parallelism}
Language extensions and DSLs aim to combat the expressivity problems of HLS.
They extend C with CSP-like parallelism~\cite{handelc},
exploit software-oriented parallel interfaces in C\#~\cite{kiwi},
or start with SystemC instead of plain C~\cite{panda:systemc, catapulthls}.
Spatial~\cite{spatial} provides primitives to generate hardware from
\emph{parallel patterns}~\cite{prabhakar:spatial-lang}.
HeteroCL~\cite{heterocl} is a Python-based DSL for optimizing programs above the HLS level of abstraction.
These languages are higher level than \sys{} and are not appropriate as
general IRs because they are tied to specific models of parallelism.
\sys{} can serve as a backend for them.

\paragraph{IRs for HDLs}
Modern HDL toolchains have IRs for transforming hardware designs~\cite{coreir, llhd, firrtl, lnast, rtlil}.
These IRs work at the RTL level of abstraction and are appropriate for representing a finished hardware implementation.
For generating and optimizing accelerators from DSLs, however, they have the same abstraction gap problem as any other RTL language.
These IRs are potential compilation targets for \sys{}.

\section{Future Work}
\label{sec:future-work}

\sys{} provides a useful foundation for exploring the design of higher-level
DSLs, compiler optimizations, and target-specific hardware design.
We plan to build upon it to explore these ideas.

\paragraph{First-class pipelining}
Pipelines are a crucial building block for high-performance hardware designs.
\sys{} program encode pipelines using \code|while| loops and \code|par|
blocks.
However, in keeping with \sys{}'s philosophy of explicit control flow, we plan to
design a first-class operator that will enable frontends to explicitly instantiate
pipelines.
An explicit representation will enable the compiler to implement
pipeline-specific optimizations such as automatic buffer insertion.
Higher-level control operators, such as pipelining, can be compiled into
more primitive control operators, which lets the \sys{} IL and compiler
incrementally add support for new operators.

\paragraph{Target-specific optimization}
\sys{}'s optimization passes do not currently use cost models and other
heuristics.
We plan to extend the \sys{} compiler to support target-specific heuristics
that enable users to make different trade-offs for different targets.
For example, multiplexers are cheap in ASICs but expensive in FPGAs while
registers are the opposite.
Such differences should affect how aggressively optimization passes that
share registers are applied.

\paragraph{Burden of synthesizability}
Several factors affect the ability of a design to meet a specific clock
period: the fan-out and fan-in factors of modules, the size of the control FSM,
and placement of registers in long combinational paths.
Currently, \sys{} requires frontends to account for these problems and generate
programs that, for example, break up long combinational paths.
In the future, we plan to implement passes that can analyze programs
for such problems and transform them to make them synthesizable.
Compiler developers can then use these passes and shift the burden of
synthesizability onto the \sys{} compiler.

\section{Conclusion}

The world of specialized hardware accelerator generators needs more shared infrastructure.
A common representation of control and structure can enable interoperability between languages while amplifying the impact of cross-cutting optimizations, analyses, transformations, and tools.

\begin{acks}
We thank Theodore Bauer and Kenneth Fang for their contributions to the
implementation of the \sys{} compiler.
Drew Zagieboylo and Zhiru Zhang provided feedback on the design of \sys{} and
early drafts of the paper.
Luis Vega provided invaluable help in understanding synthesis toolchains
debugging RTL code generation.
We also thank the anonymous reviewers and our shepherd, Sophia Shao, for their
detailed feedback.

This work was supported in part by the Center for Applications Driving Architectures (ADA), one of six centers of JUMP, a Semiconductor Research Corporation program co-sponsored by DARPA.
This is also partially supported by the Intel and NSF joint research center for Computer Assisted Programming for Heterogeneous Architectures (CAPA).
We also gratefully acknowledge support from SambaNova Systems and software donations from Xilinx.
Support included NSF awards \#1845952 and \#1723715.
\end{acks}

\appendix
\section{The \sys{} Artifact}

\subsection{Abstract}

Our artifact packages an environment that can be used to reproduce the figures
in the paper and perform similar evaluations.
It is available at the following link:

\begin{center}
\url{https://zenodo.org/record/4432747}
\end{center}

It includes the following:
\begin{itemize}
\item \code{futil}: The \sys{} compiler.
\item \code{fud}: Driver for the \code{futil} compiler and hardware tools.
\item Linear algebra PolyBench written in Dahlia.
\end{itemize}

\paragraph{Note on proprietary tools}
We use Xilinx's Vivado and Vivado HLS tools to synthesize hardware designs and
to generate HLS estimates.
While trail version of these tools can be installed using Xilinx's HL WebPACK
installer, their licenses for these tool disallow redistribution.
Our \code{README.md} details installation steps for these tools.

\subsection{Artifact check-list (meta-information)}

{\small
\begin{itemize}
  \item {\bf Program: } Polybench Benchmark Suite \cite{polybench}. (All
    benchmarks used in the evaluation are included with the artifact.)
  \item {\bf Binary: } All binaries included except Vivado and Vivado HLS.
  \item {\bf Run-time environment: } Rust source code can be compiled anywhere:
macOS, Windows, and Linux will all work. Our evaluation scripts assume a Unix
environment with the following installed:
    \begin{itemize}
    \item \code{GNU Parallel 20161222}
    \item \code{verilator v4.038}
    \item \code{python3, pip3} and the python packages: \code{numpy, pandas, seaborn, matplotlib, jupyterlab}
    \item \code{jq 1.5.1}
    \item \code{vcdump 0.1.2}
    \item \code{vivado v2019.2, vivado_hls v2019.2}
    \item \code{futil, fud} from commit \href{https://github.com/cucapra/calyx/tree/dccd6fc08ff5ed5ad38637d29610fe8ebda14354}{\code{dccd6f}}.
    \item \code{dahlia} from commit \href{https://github.com/cucapra/dahlia/tree/978ffa21572957c85a9409d80850af91b42fdaa0}{\code{978ffa}}.
    \end{itemize}
    Our packaged virtual machine has these tools installed.
  \item {\bf Metrics: } LUT usage and simulated cycle counts.
  \item {\bf Output: } The figures reported in the paper.
  \item {\bf Experiments: } We provide scripts for running the experiments and
    use Jupyter notebook for making the figures.
  \item {\bf How much disk space required (approximately)?: } 65~GB.
  \item {\bf Time needed to prepare workflow?: } 4--8 hours.
  \item {\bf Time needed to complete experiments?: } 4--8 hours.
\end{itemize}

\subsection{Description and Installation}

\subsubsection{How to Access}

The artifact is provided in two forms:
\begin{itemize}
\item A virtual image with all dependencies installed.

\item Code repositories hosted on GitHub.
\end{itemize}
The instructions to download both the virtual image and the code repositories
can be accessed here:

\begin{center}
\url{https://github.com/cucapra/calyx-evaluation}
\end{center}

To install the proprietary tools and run the scripts, please follow the instructions in the \code{README.md} file at the root of the code repository.

\subsection{Evaluation and Expected Results}

The evaluation process aims to accomplish two goals:
\begin{itemize}
\item Reproduce the graphs in the paper (Figures 5 and 6).
\item Demonstrate the robustness of our tooling and infrastructure.
\end{itemize}
The \code{README.md} file at the root of the code repository walks through
the steps to reproduce the graphs from the paper,
use the compiler to generate RTL code,
and build on the infrastructure as a library.

\paragraph{Note on \cref{fig:systolic-eval:cycles}}
Our original submission contained a bug in one of the plotting scripts that
was caught and fixed during artifact evaluation process.
Complete details are in the \code{README.md} instructions.

\subsection{Methodology}

\href{http://cTuning.org/ae/submission-20201122.html}{Submission},
\href{http://cTuning.org/ae/reviewing-20201122.html}{reviewing}, and
\href{https://www.acm.org/publications/policies/artifact-review-badging}{badging methodology}.

\bibliography{./bib/venues,./bib/papers}


\begin{thebibliography}{45}


\ifx \showCODEN    \undefined \def \showCODEN     #1{\unskip}     \fi
\ifx \showDOI      \undefined \def \showDOI       #1{#1}\fi
\ifx \showISBNx    \undefined \def \showISBNx     #1{\unskip}     \fi
\ifx \showISBNxiii \undefined \def \showISBNxiii  #1{\unskip}     \fi
\ifx \showISSN     \undefined \def \showISSN      #1{\unskip}     \fi
\ifx \showLCCN     \undefined \def \showLCCN      #1{\unskip}     \fi
\ifx \shownote     \undefined \def \shownote      #1{#1}          \fi
\ifx \showarticletitle \undefined \def \showarticletitle #1{#1}   \fi
\ifx \showURL      \undefined \def \showURL       {\relax}        \fi
\providecommand\bibfield[2]{#2}
\providecommand\bibinfo[2]{#2}
\providecommand\natexlab[1]{#1}
\providecommand\showeprint[2][]{arXiv:#2}

\bibitem[\protect\citeauthoryear{Abdallah and Hawkins}{Abdallah and
  Hawkins}{2003}]%
        {handelc}
\bibfield{author}{\bibinfo{person}{Ali~E. Abdallah} {and} \bibinfo{person}{John
  Hawkins}.} \bibinfo{year}{2003}\natexlab{}.
\newblock \showarticletitle{Formal Behavioural Synthesis of {Handel-C} Parallel
  Hardware Implementations from Functional Specifications}. In
  \bibinfo{booktitle}{\emph{Hawaii International Conference on System Sciences
  (HICSS)}}.
\newblock


\bibitem[\protect\citeauthoryear{Ananian}{Ananian}{1998}]%
        {suifhls}
\bibfield{author}{\bibinfo{person}{C~Scott Ananian}.}
  \bibinfo{year}{1998}\natexlab{}.
\newblock \bibinfo{title}{{Silicon C}: A Hardware Backend for {SUIF}}.
\newblock
\newblock
\newblock
\shownote{\url{https://flex.cscott.net/SiliconC/}.}


\bibitem[\protect\citeauthoryear{Canis, Choi, Aldham, Zhang, Kammoona,
  Anderson, Brown, and Czajkowski}{Canis et~al\mbox{.}}{2011}]%
        {legup}
\bibfield{author}{\bibinfo{person}{Andrew Canis}, \bibinfo{person}{Jongsok
  Choi}, \bibinfo{person}{Mark Aldham}, \bibinfo{person}{Victor Zhang},
  \bibinfo{person}{Ahmed Kammoona}, \bibinfo{person}{Jason~H Anderson},
  \bibinfo{person}{Stephen Brown}, {and} \bibinfo{person}{Tomasz Czajkowski}.}
  \bibinfo{year}{2011}\natexlab{}.
\newblock \showarticletitle{{LegUp}: High-level synthesis for {FPGA}-based
  processor/accelerator systems}. In \bibinfo{booktitle}{\emph{International
  Symposium on Field-Programmable Gate Arrays (FPGA)}}.
\newblock


\bibitem[\protect\citeauthoryear{Carloni, McMillan, and
  Sangiovanni-Vincentelli}{Carloni et~al\mbox{.}}{2001}]%
        {carloni:latency-insensitive}
\bibfield{author}{\bibinfo{person}{Luca~P Carloni}, \bibinfo{person}{Kenneth~L
  McMillan}, {and} \bibinfo{person}{Alberto~L Sangiovanni-Vincentelli}.}
  \bibinfo{year}{2001}\natexlab{}.
\newblock \showarticletitle{Theory of latency-insensitive design}.
\newblock \bibinfo{journal}{\emph{IEEE/ACM International Conference on
  Computer-Aided Design (ICCAD)}} (\bibinfo{year}{2001}).
\newblock


\bibitem[\protect\citeauthoryear{{Cong} and {Wang}}{{Cong} and {Wang}}{2018}]%
        {polysa}
\bibfield{author}{\bibinfo{person}{J. {Cong}} {and} \bibinfo{person}{J.
  {Wang}}.} \bibinfo{year}{2018}\natexlab{}.
\newblock \showarticletitle{{PolySA}: Polyhedral-Based Systolic Array
  Auto-Compilation}. In \bibinfo{booktitle}{\emph{IEEE/ACM International
  Conference on Computer-Aided Design (ICCAD)}}.
\newblock


\bibitem[\protect\citeauthoryear{Cong and Zhang}{Cong and Zhang}{2006}]%
        {sdc-sched}
\bibfield{author}{\bibinfo{person}{J. Cong} {and} \bibinfo{person}{Zhiru
  Zhang}.} \bibinfo{year}{2006}\natexlab{}.
\newblock \showarticletitle{An efficient and versatile scheduling algorithm
  based on {SDC} formulation}. In \bibinfo{booktitle}{\emph{Design Automation
  Conference (DAC)}}.
\newblock


\bibitem[\protect\citeauthoryear{Daly, Truong, and Hanrahan}{Daly
  et~al\mbox{.}}{2018}]%
        {coreir}
\bibfield{author}{\bibinfo{person}{Ross Daly}, \bibinfo{person}{Lenny Truong},
  {and} \bibinfo{person}{Pat Hanrahan}.} \bibinfo{year}{2018}\natexlab{}.
\newblock \showarticletitle{Invoking and Linking Generators from Multiple
  Hardware Languages using {CoreIR}}. In \bibinfo{booktitle}{\emph{Second
  Workshop on Open-Source EDA Technology (WOSET)}}.
\newblock


\bibitem[\protect\citeauthoryear{Durst, Feldman, Huff, Akeley, Daly, Bernstein,
  Patrignani, Fatahalian, and Hanrahan}{Durst et~al\mbox{.}}{2020}]%
        {aetherling}
\bibfield{author}{\bibinfo{person}{David Durst}, \bibinfo{person}{Matthew
  Feldman}, \bibinfo{person}{Dillon Huff}, \bibinfo{person}{David Akeley},
  \bibinfo{person}{Ross Daly}, \bibinfo{person}{Gilbert~Louis Bernstein},
  \bibinfo{person}{Marco Patrignani}, \bibinfo{person}{Kayvon Fatahalian},
  {and} \bibinfo{person}{Pat Hanrahan}.} \bibinfo{year}{2020}\natexlab{}.
\newblock \showarticletitle{Type-Directed Scheduling of Streaming
  Accelerators}. In \bibinfo{booktitle}{\emph{ACM SIGPLAN Conference on
  Programming Language Design and Implementation (PLDI)}}.
\newblock


\bibitem[\protect\citeauthoryear{Dutt, Hadley, and Gajski}{Dutt
  et~al\mbox{.}}{1991}]%
        {dutt:intermediate}
\bibfield{author}{\bibinfo{person}{Nikil~D Dutt}, \bibinfo{person}{Tedd
  Hadley}, {and} \bibinfo{person}{Daniel~D Gajski}.}
  \bibinfo{year}{1991}\natexlab{}.
\newblock \showarticletitle{An intermediate representation for behavioral
  synthesis}. In \bibinfo{booktitle}{\emph{Design Automation Conference
  (DAC)}}.
\newblock


\bibitem[\protect\citeauthoryear{Fowers, Ovtcharov, Papamichael, Massengill,
  Liu, Lo, Alkalay, Haselman, Adams, Ghandi, Heil, Patel, Sapek, Weisz, Woods,
  Lanka, Reinhardt, Caulfield, Chung, and Burger}{Fowers et~al\mbox{.}}{2018}]%
        {brainwave}
\bibfield{author}{\bibinfo{person}{Jeremy Fowers}, \bibinfo{person}{Kalin
  Ovtcharov}, \bibinfo{person}{Michael Papamichael}, \bibinfo{person}{Todd
  Massengill}, \bibinfo{person}{Ming Liu}, \bibinfo{person}{Daniel Lo},
  \bibinfo{person}{Shlomi Alkalay}, \bibinfo{person}{Michael Haselman},
  \bibinfo{person}{Logan Adams}, \bibinfo{person}{Mahdi Ghandi},
  \bibinfo{person}{Stephen Heil}, \bibinfo{person}{Prerak Patel},
  \bibinfo{person}{Adam Sapek}, \bibinfo{person}{Gabriel Weisz},
  \bibinfo{person}{Lisa Woods}, \bibinfo{person}{Sitaram Lanka},
  \bibinfo{person}{Steven~K. Reinhardt}, \bibinfo{person}{Adrian~M. Caulfield},
  \bibinfo{person}{Eric~S. Chung}, {and} \bibinfo{person}{Doug Burger}.}
  \bibinfo{year}{2018}\natexlab{}.
\newblock \showarticletitle{A Configurable Cloud-scale {DNN} Processor for
  Real-time {AI}}. In \bibinfo{booktitle}{\emph{International Symposium on
  Computer Architecture (ISCA)}}.
\newblock


\bibitem[\protect\citeauthoryear{Guo, Buyukkurt, Cortes, Mitra, and Najjar}{Guo
  et~al\mbox{.}}{2008}]%
        {guo:cirrf}
\bibfield{author}{\bibinfo{person}{Zhi Guo}, \bibinfo{person}{Betul Buyukkurt},
  \bibinfo{person}{John Cortes}, \bibinfo{person}{Abhishek Mitra}, {and}
  \bibinfo{person}{Walild Najjar}.} \bibinfo{year}{2008}\natexlab{}.
\newblock \showarticletitle{A compiler intermediate representation for
  reconfigurable fabrics}.
\newblock \bibinfo{journal}{\emph{International Journal of Parallel
  Programming}} (\bibinfo{year}{2008}).
\newblock


\bibitem[\protect\citeauthoryear{Gupta, Gupta, Dutt, and Nicolau}{Gupta
  et~al\mbox{.}}{2004}]%
        {gupta:spark}
\bibfield{author}{\bibinfo{person}{S Gupta}, \bibinfo{person}{Renu Gupta},
  \bibinfo{person}{Nikil Dutt}, {and} \bibinfo{person}{Alex Nicolau}.}
  \bibinfo{year}{2004}\natexlab{}.
\newblock \bibinfo{booktitle}{\emph{{SPARK}: A Parallelizing Approach to the
  High-Level Synthesis of Digital Circuits}}.
\newblock


\bibitem[\protect\citeauthoryear{Hegarty, Brunhaver, DeVito, Ragan-Kelley,
  Cohen, Bell, Vasilyev, Horowitz, and Hanrahan}{Hegarty et~al\mbox{.}}{2014}]%
        {hegarty:darkroom}
\bibfield{author}{\bibinfo{person}{James Hegarty}, \bibinfo{person}{John
  Brunhaver}, \bibinfo{person}{Zachary DeVito}, \bibinfo{person}{Jonathan
  Ragan-Kelley}, \bibinfo{person}{Noy Cohen}, \bibinfo{person}{Steven Bell},
  \bibinfo{person}{Artem Vasilyev}, \bibinfo{person}{Mark Horowitz}, {and}
  \bibinfo{person}{Pat Hanrahan}.} \bibinfo{year}{2014}\natexlab{}.
\newblock \showarticletitle{{Darkroom}: Compiling high-level image processing
  code into hardware pipelines}.
\newblock \bibinfo{journal}{\emph{ACM Transactions on Graphics}}.
\newblock


\bibitem[\protect\citeauthoryear{{Intel}}{{Intel}}{2021}]%
        {intelhls}
\bibfield{author}{\bibinfo{person}{{Intel}}.} \bibinfo{year}{2021}\natexlab{}.
\newblock \bibinfo{booktitle}{\emph{{Intel High Level Synthesis Compiler}}}.
\newblock
\urldef\tempurl%
\url{https://www.altera.com/products/design-software/high-level-design/intel-hls-compiler/overview.html}
\showURL{%
Retrieved January 16, 2021 from \tempurl}


\bibitem[\protect\citeauthoryear{Izraelevitz, Koenig, Li, Lin, Wang, Magyar,
  Kim, Schmidt, Markley, Lawson, and Bachrach}{Izraelevitz
  et~al\mbox{.}}{2017}]%
        {firrtl}
\bibfield{author}{\bibinfo{person}{Adam~M. Izraelevitz}, \bibinfo{person}{Jack
  Koenig}, \bibinfo{person}{Patrick Li}, \bibinfo{person}{Richard Lin},
  \bibinfo{person}{Angie Wang}, \bibinfo{person}{Albert Magyar},
  \bibinfo{person}{Donggyu Kim}, \bibinfo{person}{Colin Schmidt},
  \bibinfo{person}{Chick Markley}, \bibinfo{person}{Jim Lawson}, {and}
  \bibinfo{person}{Jonathan Bachrach}.} \bibinfo{year}{2017}\natexlab{}.
\newblock \showarticletitle{Reusability is {FIRRTL} ground: Hardware
  construction languages, compiler frameworks, and transformations}. In
  \bibinfo{booktitle}{\emph{IEEE/ACM International Conference on Computer-Aided
  Design (ICCAD)}}.
\newblock


\bibitem[\protect\citeauthoryear{Josipoviundefined, Ghosal, and
  Ienne}{Josipoviundefined et~al\mbox{.}}{2018}]%
        {dynamic-schedule-hls}
\bibfield{author}{\bibinfo{person}{Lana Josipoviundefined},
  \bibinfo{person}{Radhika Ghosal}, {and} \bibinfo{person}{Paolo Ienne}.}
  \bibinfo{year}{2018}\natexlab{}.
\newblock \showarticletitle{Dynamically Scheduled High-Level Synthesis}. In
  \bibinfo{booktitle}{\emph{International Symposium on Field-Programmable Gate
  Arrays (FPGA)}}.
\newblock


\bibitem[\protect\citeauthoryear{Jouppi, Young, Patil, Patterson, Agrawal,
  Bajwa, Bates, Bhatia, Boden, Borchers, Boyle, luc Cantin, Chao, Clark,
  Coriell, Daley, Dau, Dean, Gelb, Ghaemmaghami, Gottipati, Gulland, Hagmann,
  Ho, Hogberg, Hu, Hundt, Hurt, Ibarz, Jaffey, Jaworski, Kaplan, Khaitan, Koch,
  Kumar, Lacy, Laudon, Law, Le, Leary, Liu, Lucke, Lundin, MacKean, Maggiore,
  Mahony, Miller, Nagarajan, Narayanaswami, Ni, Nix, Norrie, Omernick,
  Penukonda, Phelps, Ross, Ross, Salek, Samadiani, Severn, Sizikov, Snelham,
  Souter, Steinberg, Swing, Tan, Thorson, Tian, Toma, Tuttle, Vasudevan,
  Walter, Wang, Wilcox, and Yoon}{Jouppi et~al\mbox{.}}{2017}]%
        {tpu}
\bibfield{author}{\bibinfo{person}{Norman~P. Jouppi}, \bibinfo{person}{Cliff
  Young}, \bibinfo{person}{Nishant Patil}, \bibinfo{person}{David Patterson},
  \bibinfo{person}{Gaurav Agrawal}, \bibinfo{person}{Raminder Bajwa},
  \bibinfo{person}{Sarah Bates}, \bibinfo{person}{Suresh Bhatia},
  \bibinfo{person}{Nan Boden}, \bibinfo{person}{Al Borchers},
  \bibinfo{person}{Rick Boyle}, \bibinfo{person}{Pierre luc Cantin},
  \bibinfo{person}{Clifford Chao}, \bibinfo{person}{Chris Clark},
  \bibinfo{person}{Jeremy Coriell}, \bibinfo{person}{Mike Daley},
  \bibinfo{person}{Matt Dau}, \bibinfo{person}{Jeffrey Dean},
  \bibinfo{person}{Ben Gelb}, \bibinfo{person}{Tara~Vazir Ghaemmaghami},
  \bibinfo{person}{Rajendra Gottipati}, \bibinfo{person}{William Gulland},
  \bibinfo{person}{Robert Hagmann}, \bibinfo{person}{C.~Richard Ho},
  \bibinfo{person}{Doug Hogberg}, \bibinfo{person}{John Hu},
  \bibinfo{person}{Robert Hundt}, \bibinfo{person}{Dan Hurt},
  \bibinfo{person}{Julian Ibarz}, \bibinfo{person}{Aaron Jaffey},
  \bibinfo{person}{Alek Jaworski}, \bibinfo{person}{Alexander Kaplan},
  \bibinfo{person}{Harshit Khaitan}, \bibinfo{person}{Andy Koch},
  \bibinfo{person}{Naveen Kumar}, \bibinfo{person}{Steve Lacy},
  \bibinfo{person}{James Laudon}, \bibinfo{person}{James Law},
  \bibinfo{person}{Diemthu Le}, \bibinfo{person}{Chris Leary},
  \bibinfo{person}{Zhuyuan Liu}, \bibinfo{person}{Kyle Lucke},
  \bibinfo{person}{Alan Lundin}, \bibinfo{person}{Gordon MacKean},
  \bibinfo{person}{Adriana Maggiore}, \bibinfo{person}{Maire Mahony},
  \bibinfo{person}{Kieran Miller}, \bibinfo{person}{Rahul Nagarajan},
  \bibinfo{person}{Ravi Narayanaswami}, \bibinfo{person}{Ray Ni},
  \bibinfo{person}{Kathy Nix}, \bibinfo{person}{Thomas Norrie},
  \bibinfo{person}{Mark Omernick}, \bibinfo{person}{Narayana Penukonda},
  \bibinfo{person}{Andy Phelps}, \bibinfo{person}{Jonathan Ross},
  \bibinfo{person}{Matt Ross}, \bibinfo{person}{Amir Salek},
  \bibinfo{person}{Emad Samadiani}, \bibinfo{person}{Chris Severn},
  \bibinfo{person}{Gregory Sizikov}, \bibinfo{person}{Matthew Snelham},
  \bibinfo{person}{Jed Souter}, \bibinfo{person}{Dan Steinberg},
  \bibinfo{person}{Andy Swing}, \bibinfo{person}{Mercedes Tan},
  \bibinfo{person}{Gregory Thorson}, \bibinfo{person}{Bo Tian},
  \bibinfo{person}{Horia Toma}, \bibinfo{person}{Erick Tuttle},
  \bibinfo{person}{Vijay Vasudevan}, \bibinfo{person}{Richard Walter},
  \bibinfo{person}{Walter Wang}, \bibinfo{person}{Eric Wilcox}, {and}
  \bibinfo{person}{Doe~Hyun Yoon}.} \bibinfo{year}{2017}\natexlab{}.
\newblock \showarticletitle{In-Datacenter Performance Analysis of a {Tensor
  Processing Unit}}. In \bibinfo{booktitle}{\emph{International Symposium on
  Computer Architecture (ISCA)}}.
\newblock


\bibitem[\protect\citeauthoryear{Koeplinger, Feldman, Prabhakar, Zhang, Hadjis,
  Fiszel, Zhao, Nardi, Pedram, Kozyrakis, and Olukotun}{Koeplinger
  et~al\mbox{.}}{2018}]%
        {spatial}
\bibfield{author}{\bibinfo{person}{David Koeplinger}, \bibinfo{person}{Matthew
  Feldman}, \bibinfo{person}{Raghu Prabhakar}, \bibinfo{person}{Yaqi Zhang},
  \bibinfo{person}{Stefan Hadjis}, \bibinfo{person}{Ruben Fiszel},
  \bibinfo{person}{Tian Zhao}, \bibinfo{person}{Luigi Nardi},
  \bibinfo{person}{Ardavan Pedram}, \bibinfo{person}{Christos Kozyrakis}, {and}
  \bibinfo{person}{Kunle Olukotun}.} \bibinfo{year}{2018}\natexlab{}.
\newblock \showarticletitle{Spatial: A language and compiler for application
  accelerators}. In \bibinfo{booktitle}{\emph{ACM SIGPLAN Conference on
  Programming Language Design and Implementation (PLDI)}}.
\newblock


\bibitem[\protect\citeauthoryear{Kung}{Kung}{1982}]%
        {kung:systolic}
\bibfield{author}{\bibinfo{person}{Hsiang-Tsung Kung}.}
  \bibinfo{year}{1982}\natexlab{}.
\newblock \showarticletitle{Why systolic architectures?}
\newblock \bibinfo{journal}{\emph{IEEE computer}} (\bibinfo{year}{1982}).
\newblock


\bibitem[\protect\citeauthoryear{Lai, Chi, Hu, Wang, Yu, Zhou, Cong, and
  Zhang}{Lai et~al\mbox{.}}{2019}]%
        {heterocl}
\bibfield{author}{\bibinfo{person}{Yi-Hsiang Lai}, \bibinfo{person}{Yuze Chi},
  \bibinfo{person}{Yuwei Hu}, \bibinfo{person}{Jie Wang},
  \bibinfo{person}{Cody~Hao Yu}, \bibinfo{person}{Yuan Zhou},
  \bibinfo{person}{Jason Cong}, {and} \bibinfo{person}{Zhiru Zhang}.}
  \bibinfo{year}{2019}\natexlab{}.
\newblock \showarticletitle{{HeteroCL}: A Multi-Paradigm Programming
  Infrastructure for Software-Defined Reconfigurable Computing}. In
  \bibinfo{booktitle}{\emph{International Symposium on Field-Programmable Gate
  Arrays (FPGA)}}.
\newblock


\bibitem[\protect\citeauthoryear{Lai, Rong, Zheng, Zhang, Cui, Jia, Wang,
  Sullivan, Zhang, Liang, Zhang, Cong, George, Alvarez, Hughes, and Dubey}{Lai
  et~al\mbox{.}}{2020}]%
        {susy}
\bibfield{author}{\bibinfo{person}{Y.-H. Lai}, \bibinfo{person}{H. Rong},
  \bibinfo{person}{S. Zheng}, \bibinfo{person}{W. Zhang}, \bibinfo{person}{X.
  Cui}, \bibinfo{person}{Y. Jia}, \bibinfo{person}{J. Wang},
  \bibinfo{person}{B. Sullivan}, \bibinfo{person}{Z. Zhang},
  \bibinfo{person}{Y. Liang}, \bibinfo{person}{Y. Zhang}, \bibinfo{person}{J.
  Cong}, \bibinfo{person}{N. George}, \bibinfo{person}{J. Alvarez},
  \bibinfo{person}{C. Hughes}, {and} \bibinfo{person}{P. Dubey}.}
  \bibinfo{year}{2020}\natexlab{}.
\newblock \showarticletitle{{SuSy}: A Programming Model for Productive
  Construction of High-Performance Systolic Arrays on {FPGAs}}. In
  \bibinfo{booktitle}{\emph{IEEE/ACM International Conference on Computer-Aided
  Design (ICCAD)}}.
\newblock


\bibitem[\protect\citeauthoryear{Lattner and Adve}{Lattner and Adve}{2004}]%
        {llvm}
\bibfield{author}{\bibinfo{person}{Chris Lattner} {and} \bibinfo{person}{Vikram
  Adve}.} \bibinfo{year}{2004}\natexlab{}.
\newblock \showarticletitle{{LLVM}: A Compilation Framework for Lifelong
  Program Analysis \& Transformation}. In
  \bibinfo{booktitle}{\emph{International Symposium on Code Generation and
  Optimization (CGO)}}.
\newblock


\bibitem[\protect\citeauthoryear{{Louis-Noel Pouchet}}{{Louis-Noel
  Pouchet}}{2021}]%
        {polybench}
\bibfield{author}{\bibinfo{person}{{Louis-Noel Pouchet}}.}
  \bibinfo{year}{2021}\natexlab{}.
\newblock \bibinfo{booktitle}{\emph{{PolyBench/C: The Polyhedral Benchmark
  Suite.}}}
\newblock
\urldef\tempurl%
\url{http://web.cse.ohio-state.edu/~pouchet.2/software/polybench/}
\showURL{%
Retrieved January 16, 2021 from \tempurl}


\bibitem[\protect\citeauthoryear{{Mentor Graphics}}{{Mentor Graphics}}{2021}]%
        {catapulthls}
\bibfield{author}{\bibinfo{person}{{Mentor Graphics}}.}
  \bibinfo{year}{2021}\natexlab{}.
\newblock \bibinfo{booktitle}{\emph{{Catapult} High-Level Synthesis}}.
\newblock
\urldef\tempurl%
\url{https://www.mentor.com/hls-lp/catapult-high-level-synthesis/}
\showURL{%
Retrieved January 16, 2021 from \tempurl}


\bibitem[\protect\citeauthoryear{Nigam, Atapattu, Thomas, Li, Bauer, Ye, Koti,
  Sampson, and Zhang}{Nigam et~al\mbox{.}}{2020}]%
        {dahlia}
\bibfield{author}{\bibinfo{person}{Rachit Nigam}, \bibinfo{person}{Sachille
  Atapattu}, \bibinfo{person}{Samuel Thomas}, \bibinfo{person}{Zhijing Li},
  \bibinfo{person}{Theodore Bauer}, \bibinfo{person}{Yuwei Ye},
  \bibinfo{person}{Apurva Koti}, \bibinfo{person}{Adrian Sampson}, {and}
  \bibinfo{person}{Zhiru Zhang}.} \bibinfo{year}{2020}\natexlab{}.
\newblock \showarticletitle{Predictable Accelerator Design with Time-Sensitive
  Affine Types}. In \bibinfo{booktitle}{\emph{ACM SIGPLAN Conference on
  Programming Language Design and Implementation (PLDI)}}.
\newblock


\bibitem[\protect\citeauthoryear{Nikhil}{Nikhil}{2004}]%
        {nikhil:bluespec}
\bibfield{author}{\bibinfo{person}{Rishiyur Nikhil}.}
  \bibinfo{year}{2004}\natexlab{}.
\newblock \showarticletitle{{Bluespec System Verilog}: Efficient, correct {RTL}
  from high level specifications}. In \bibinfo{booktitle}{\emph{Conference on
  Formal Methods and Models for Co-Design (MEMOCODE)}}.
\newblock


\bibitem[\protect\citeauthoryear{Panda}{Panda}{2001}]%
        {panda:systemc}
\bibfield{author}{\bibinfo{person}{Preeti~Ranjan Panda}.}
  \bibinfo{year}{2001}\natexlab{}.
\newblock \showarticletitle{{SystemC}: A modeling platform supporting multiple
  design abstractions}. In \bibinfo{booktitle}{\emph{International Symposium on
  Systems Synthesis}}.
\newblock


\bibitem[\protect\citeauthoryear{Pilato and Ferrandi}{Pilato and
  Ferrandi}{2013}]%
        {bambu}
\bibfield{author}{\bibinfo{person}{Christian Pilato} {and}
  \bibinfo{person}{Fabrizio Ferrandi}.} \bibinfo{year}{2013}\natexlab{}.
\newblock \showarticletitle{{Bambu}: A modular framework for the high level
  synthesis of memory-intensive applications}. In
  \bibinfo{booktitle}{\emph{International Conference on Field-Programmable
  Logic and Applications (FPL)}}.
\newblock


\bibitem[\protect\citeauthoryear{Prabhakar, Koeplinger, Brown, Lee, De~Sa,
  Kozyrakis, and Olukotun}{Prabhakar et~al\mbox{.}}{2016}]%
        {prabhakar:spatial-lang}
\bibfield{author}{\bibinfo{person}{Raghu Prabhakar}, \bibinfo{person}{David
  Koeplinger}, \bibinfo{person}{Kevin~J Brown}, \bibinfo{person}{HyoukJoong
  Lee}, \bibinfo{person}{Christopher De~Sa}, \bibinfo{person}{Christos
  Kozyrakis}, {and} \bibinfo{person}{Kunle Olukotun}.}
  \bibinfo{year}{2016}\natexlab{}.
\newblock \showarticletitle{Generating configurable hardware from parallel
  patterns}. In \bibinfo{booktitle}{\emph{ACM International Conference on
  Architectural Support for Programming Languages and Operating Systems
  (ASPLOS)}}.
\newblock


\bibitem[\protect\citeauthoryear{Pu, Bell, Yang, Setter, Richardson,
  Ragan-Kelley, and Horowitz}{Pu et~al\mbox{.}}{2017}]%
        {halide-hls}
\bibfield{author}{\bibinfo{person}{Jing Pu}, \bibinfo{person}{Steven Bell},
  \bibinfo{person}{Xuan Yang}, \bibinfo{person}{Jeff Setter},
  \bibinfo{person}{Stephen Richardson}, \bibinfo{person}{Jonathan
  Ragan-Kelley}, {and} \bibinfo{person}{Mark Horowitz}.}
  \bibinfo{year}{2017}\natexlab{}.
\newblock \showarticletitle{Programming heterogeneous systems from an image
  processing {DSL}}.
\newblock \bibinfo{journal}{\emph{ACM Transactions on Architecture and Code
  Optimization (TACO)}}.
\newblock


\bibitem[\protect\citeauthoryear{Ragan{-}Kelley, Barnes, Adams, Paris, Durand,
  and Amarasinghe}{Ragan{-}Kelley et~al\mbox{.}}{2013}]%
        {ragan-kelley:halide}
\bibfield{author}{\bibinfo{person}{Jonathan Ragan{-}Kelley},
  \bibinfo{person}{Connelly Barnes}, \bibinfo{person}{Andrew Adams},
  \bibinfo{person}{Sylvain Paris}, \bibinfo{person}{Fr{\'{e}}do Durand}, {and}
  \bibinfo{person}{Saman~P. Amarasinghe}.} \bibinfo{year}{2013}\natexlab{}.
\newblock \showarticletitle{Halide: A language and compiler for optimizing
  parallelism, locality, and recomputation in image processing pipelines}. In
  \bibinfo{booktitle}{\emph{ACM SIGPLAN Conference on Programming Language
  Design and Implementation (PLDI)}}.
\newblock


\bibitem[\protect\citeauthoryear{Sahasrabuddhe, Raja, Arya, and
  Desai}{Sahasrabuddhe et~al\mbox{.}}{2007}]%
        {ahir}
\bibfield{author}{\bibinfo{person}{Sameer~D Sahasrabuddhe},
  \bibinfo{person}{Hakim Raja}, \bibinfo{person}{Kavi Arya}, {and}
  \bibinfo{person}{Madhav~P Desai}.} \bibinfo{year}{2007}\natexlab{}.
\newblock \showarticletitle{{AHIR}: A hardware intermediate representation for
  hardware generation from high-level programs}. In
  \bibinfo{booktitle}{\emph{International Conference on VLSI Design (VLSID)}}.
\newblock


\bibitem[\protect\citeauthoryear{Schuiki, Kurth, Grosser, and Benini}{Schuiki
  et~al\mbox{.}}{2020}]%
        {llhd}
\bibfield{author}{\bibinfo{person}{Fabian Schuiki}, \bibinfo{person}{Andreas
  Kurth}, \bibinfo{person}{Tobias Grosser}, {and} \bibinfo{person}{Luca
  Benini}.} \bibinfo{year}{2020}\natexlab{}.
\newblock \showarticletitle{{LLHD}: A Multi-Level Intermediate Representation
  for Hardware Description Languages}. In \bibinfo{booktitle}{\emph{ACM SIGPLAN
  Conference on Programming Language Design and Implementation (PLDI)}}.
\newblock


\bibitem[\protect\citeauthoryear{{Shang HLS Authors}}{{Shang HLS
  Authors}}{2021}]%
        {shanghls}
\bibfield{author}{\bibinfo{person}{{Shang HLS Authors}}.}
  \bibinfo{year}{2021}\natexlab{}.
\newblock \bibinfo{booktitle}{\emph{{The Shang High-Level Synthesis
  Framework}}}.
\newblock
\urldef\tempurl%
\url{https://web.archive.org/web/20180610233052/https://github.com/etherzhhb/Shang}
\showURL{%
Retrieved January 16, 2021 from \tempurl}


\bibitem[\protect\citeauthoryear{Sharifian, Hojabr, Rahimi, Liu, Guha,
  Nowatzki, and Shriraman}{Sharifian et~al\mbox{.}}{2019}]%
        {muir}
\bibfield{author}{\bibinfo{person}{Amirali Sharifian}, \bibinfo{person}{Reza
  Hojabr}, \bibinfo{person}{Navid Rahimi}, \bibinfo{person}{Sihao Liu},
  \bibinfo{person}{Apala Guha}, \bibinfo{person}{Tony Nowatzki}, {and}
  \bibinfo{person}{Arrvindh Shriraman}.} \bibinfo{year}{2019}\natexlab{}.
\newblock \showarticletitle{$\mu$IR: An Intermediate Representation for
  Transforming and Optimizing the Microarchitecture of Application
  Accelerators}. In \bibinfo{booktitle}{\emph{{IEEE/ACM} International
  Symposium on Microarchitecture (MICRO)}}.
\newblock


\bibitem[\protect\citeauthoryear{Singh and Greaves}{Singh and Greaves}{2008}]%
        {kiwi}
\bibfield{author}{\bibinfo{person}{Satnam Singh} {and}
  \bibinfo{person}{David~J. Greaves}.} \bibinfo{year}{2008}\natexlab{}.
\newblock \showarticletitle{{Kiwi}: Synthesis of {FPGA} Circuits from Parallel
  Programs}. In \bibinfo{booktitle}{\emph{Field-Programmable Custom Computing
  Machines (FCCM)}}.
\newblock


\bibitem[\protect\citeauthoryear{Sinha and Patel}{Sinha and Patel}{2012}]%
        {synasm}
\bibfield{author}{\bibinfo{person}{Rohit Sinha} {and} \bibinfo{person}{Hiren~D
  Patel}.} \bibinfo{year}{2012}\natexlab{}.
\newblock \showarticletitle{{synASM}: A high-level synthesis framework with
  support for parallel and timed constructs}.
\newblock \bibinfo{journal}{\emph{IEEE/ACM International Conference on
  Computer-Aided Design (ICCAD)}}.
\newblock


\bibitem[\protect\citeauthoryear{Srinivasan and Wolfe}{Srinivasan and
  Wolfe}{1992}]%
        {explicit-parallelism}
\bibfield{author}{\bibinfo{person}{H. Srinivasan} {and} \bibinfo{person}{M.
  Wolfe}.} \bibinfo{year}{1992}\natexlab{}.
\newblock \showarticletitle{Analyzing programs with explicit parallelism}. In
  \bibinfo{booktitle}{\emph{Languages and Compilers for Parallel Computing}}.
\newblock


\bibitem[\protect\citeauthoryear{{Veripool}}{{Veripool}}{2021}]%
        {verilator}
\bibfield{author}{\bibinfo{person}{{Veripool}}.}
  \bibinfo{year}{2021}\natexlab{}.
\newblock \bibinfo{title}{{Verilator}}.
\newblock
\newblock
\newblock
\shownote{\url{https://www.veripool.org/wiki/verilator}.}


\bibitem[\protect\citeauthoryear{Wang, Soul\'{e}, Dang, Lee, Shrivastav,
  Foster, and Weatherspoon}{Wang et~al\mbox{.}}{2017}]%
        {p4fpga}
\bibfield{author}{\bibinfo{person}{Han Wang}, \bibinfo{person}{Robert
  Soul\'{e}}, \bibinfo{person}{Huynh~Tu Dang}, \bibinfo{person}{Ki~Suh Lee},
  \bibinfo{person}{Vishal Shrivastav}, \bibinfo{person}{Nate Foster}, {and}
  \bibinfo{person}{Hakim Weatherspoon}.} \bibinfo{year}{2017}\natexlab{}.
\newblock \showarticletitle{{P4FPGA}: A Rapid Prototyping Framework for {P4}}.
  In \bibinfo{booktitle}{\emph{Symposium on SDN Research (SOSR)}}.
\newblock


\bibitem[\protect\citeauthoryear{Wang, Sridhar, and Renau}{Wang
  et~al\mbox{.}}{2019}]%
        {lnast}
\bibfield{author}{\bibinfo{person}{Sheng-Hong Wang}, \bibinfo{person}{Akash
  Sridhar}, {and} \bibinfo{person}{Jose Renau}.}
  \bibinfo{year}{2019}\natexlab{}.
\newblock \showarticletitle{{LNAST}: A language neutral intermediate
  representation for hardware description languages}. In
  \bibinfo{booktitle}{\emph{Second Workshop on Open-Source EDA Technology
  (WOSET)}}.
\newblock


\bibitem[\protect\citeauthoryear{Wolf}{Wolf}{2021}]%
        {rtlil}
\bibfield{author}{\bibinfo{person}{Claire Wolf}.}
  \bibinfo{year}{2021}\natexlab{}.
\newblock \bibinfo{booktitle}{\emph{Yosys Manual}}.
\newblock
\urldef\tempurl%
\url{http://www.clifford.at/yosys/files/yosys\_manual.pdf}
\showURL{%
Retrieved January 16, 2021 from \tempurl}


\bibitem[\protect\citeauthoryear{Wu, Wang, Bian, Wu, and Xue}{Wu
  et~al\mbox{.}}{2002}]%
        {wu:hierarchical}
\bibfield{author}{\bibinfo{person}{Qiang Wu}, \bibinfo{person}{Yunfeng Wang},
  \bibinfo{person}{Jinian Bian}, \bibinfo{person}{Weimin Wu}, {and}
  \bibinfo{person}{Hongxi Xue}.} \bibinfo{year}{2002}\natexlab{}.
\newblock \showarticletitle{A hierarchical {CDFG} as intermediate
  representation for hardware/software codesign}. In
  \bibinfo{booktitle}{\emph{International Conference on Communications,
  Circuits and Systems (ICCCAS)}}.
\newblock


\bibitem[\protect\citeauthoryear{{Xilinx Inc.}}{{Xilinx Inc.}}{2021}]%
        {vivadohls2017}
\bibfield{author}{\bibinfo{person}{{Xilinx Inc.}}}
  \bibinfo{year}{2021}\natexlab{}.
\newblock \bibinfo{booktitle}{\emph{{Vivado Design Suite User Guide: High-Level
  Synthesis. UG902 (v2017.2) June 7, 2017.}}}
\newblock
\urldef\tempurl%
\url{https://www.xilinx.com/support/documentation/sw\_manuals/xilinx2017\_2/ug902-vivado-high-level-synthesis.pdf}
\showURL{%
Retrieved January 16, 2021 from \tempurl}


\bibitem[\protect\citeauthoryear{Zhang, Fan, Jiang, Han, Yang, and Cong}{Zhang
  et~al\mbox{.}}{2008}]%
        {autopilot}
\bibfield{author}{\bibinfo{person}{Zhiru Zhang}, \bibinfo{person}{Yiping Fan},
  \bibinfo{person}{Wei Jiang}, \bibinfo{person}{Guoling Han},
  \bibinfo{person}{Changqi Yang}, {and} \bibinfo{person}{Jason Cong}.}
  \bibinfo{year}{2008}\natexlab{}.
\newblock \showarticletitle{{AutoPilot}: A platform-based {ESL} synthesis
  system}.
\newblock In \bibinfo{booktitle}{\emph{High-Level Synthesis}}.
  \bibinfo{pages}{99--112}.
\newblock


\end{thebibliography}

\end{document}